**UNIVERSIDADE FEDERAL DO RIO GRANDE DO SUL**

**FACULDADE DE CIÊNCIAS ECONÔMICAS**

**PROGRAMA DE PÓS-GRADUAÇÃO EM ECONOMIA**

**RAFAEL ROCKENBACH DA SILVA GUIMARÃES**

**DEEP LEARNING MACROECONOMICS**

**Porto Alegre**

**2021**

**RAFAEL ROCKENBACH DA SILVA GUIMARÃES**

**DEEP LEARNING MACROECONOMICS**

Thesis presented in partial fulfillment of the requirements for the degree of Doctor in Economics.

Supervisor: Prof. Dr. Sergio Marley Modesto Monteiro

**Porto Alegre**

**2021**

CIP - Catalogação na Publicação

**RAFAEL ROCKENBACH DA SILVA GUIMARÃES**

**DEEP LEARNING MACROECONOMICS**

> Tese submetida ao Programa de Pós-Graduação em Economia da Faculdade de Ciências Econômicas da UFRGS, como requisito parcial para obtenção do título de Doutor em Economia.

Aprovada em: Porto Alegre, 8 de novembro de 2021.

BANCA EXAMINADORA:

______________________________________________________________

Prof. Dr. Sergio Marley Modesto Monteiro – Orientador

Universidade Federal do Rio Grande do Sul – UFRGS

______________________________________________________________

Prof. Dr. Alessandro Donadio Miebach

Universidade Federal do Rio Grande do Sul - UFRGS

______________________________________________________________

Prof. Dr. Daniel Oliveira Cajueiro

Universidade de Brasília - UnB

______________________________________________________________

Prof. Dr. Wagner Piazza Gaglianone

Banco Central do Brasil

This thesis work is dedicated to Fernanda, Rafaela, and Felipe Preto Guimarães.

ACKNOWLEDGEMENTS

The view I have on the acknowledgments is those scenes at awards, like in The Oscars, where the winner has little time to talk and a massive list of people to thank. Time runs out, sound cuts out, but there are still people on the list. In this journey in search of a doctorate, countless people helped me. I will name a few that have had a more direct influence.

Firstly, my wife Fernanda and my children Rafaela and Felipe. They had to live with the day-to-day life of a doctoral student. It wasn't easy for them at times, but their support has always been unconditional. As I work at the Central Bank of Brazil, I have access to co-workers whose qualifications resulted in excellent suggestions and comments. Special thanks to Oswaldo Baumgarten Filho, friend and mentor for the deep learning topic, and Vera Maria Schneider, friend and person who identified my potential to work in the Economics Department. Thanks to Marcelo Antonio Thomaz de Aragão and Wagner Piazza Gaglianone, who had reviewed my work in internal seminars, offering excellent suggestions. I would also like to thank UniBC, which performs extraordinary work towards the qualification of the Central Bank of Brazil staff. Thanks to Professor Marcelle Chauvet for accepting my proposal to work together and welcoming me to the University of California, Riverside. Being in a foreign country and attending classes with people of different nationalities has expanded my skills. The panels' composition that evaluates the project and the thesis are crucial. Thanks to professors Alessandro Donadio Miebach, Daniel Oliveira Cajueiro, Flávio Augusto Ziegelmann, Nelson Seixas dos Santos and Wagner Piazza Gaglianone for their invaluable contributions. Finally, thanks to my advisor Professor Sérgio Marley Modesto Monteiro. We worked together for the first time during my Master's, and since then, I have been confident that he would be my first choice as a thesis advisor. As a result, I received the confidence and autonomy necessary to carry out this work. Furthermore, his guidance was instrumental in transforming my ideas into a doctoral thesis.

# RESUMO

Conjuntos de dados limitados e complexas relações não-lineares estão entre os desafios que podem surgir ao se aplicar econometria a problemas macroeconômicos. Esta pesquisa propõe a aprendizagem profunda como uma abordagem para transferir aprendizagem no primeiro caso e para mapear relações entre variáveis no último caso. Várias técnicas de aprendizado de máquina estão incorporadas à estrutura econométrica, mas o aprendizado profundo continua focado na previsão de séries temporais. A abordagem proposta aqui também está relacionada ao reconhecimento de padrões, mas onde o aprendizado profundo alcançou desempenho de ponta: progressivamente utilizando várias camadas de processamento para extrair dos dados brutos abstrações de mais alto nível.

Primeiramente, a aprendizagem por transferência é proposta como uma estratégia adicional para a macroeconomia empírica. Embora os macroeconomistas já apliquem a aprendizagem por transferência ao assumir uma dada distribuição a priori em um contexto Bayesiano, estimar um VAR estrutural com restrição de sinal e calibrar parâmetros com base em resultados observados em outros modelos, para citar alguns exemplos, avançar em uma estratégia mais sistemática de transferência de aprendizagem em macroeconomia aplicada é a inovação que estamos introduzindo. Ao desenvolver estratégias de modelagem econômica, a falta de dados pode ser um problema que a aprendizagem por transferência pode corrigir. Começamos por apresentar conceitos teóricos relacionados à transferência de aprendizagem e propomos uma conexão com uma tipologia relacionada a modelos macroeconômicos. Em seguida, exploramos a estratégia proposta empiricamente, mostrando que os dados de domínios diferentes, mas relacionados, um tipo de aprendizagem por transferência, ajudam a identificar as fases do ciclo de negócios quando não há comitê de datação do ciclo de negócios e a estimar rapidamente um hiato do produto de base econômica. Em ambos os casos, a estratégia também ajuda a melhorar o aprendizado quando os dados são limitados. A abordagem integra a ideia de armazenar conhecimento obtido de especialistas em economia de uma região e aplicá-lo a outras áreas geográficas. O primeiro é capturado com um modelo de rede neural profunda supervisionado e o segundo aplicando-o a outro conjunto de dados, um procedimento de adaptação de domínio. No geral, há uma melhora na classificação com a aprendizagem por transferência em comparação com os modelos de base. Até onde sabemos, a abordagem combinada de aprendizagem profunda e transferência é subutilizada para aplicação a problemas macroeconômicos, indicando que há muito espaço para o desenvolvimento de pesquisas.

Em segundo lugar, uma vez que os métodos de aprendizagem profunda são uma forma de aprender representações, aquelas que são formadas pela composição de várias transformações não lineares, para produzir representações mais abstratas, aplicamos a aprendizagem profunda para mapear variáveis de baixa frequência a partir de variáveis de alta frequência. Existem situações em que sabemos, às vezes por construção, que existe uma relação entre as variáveis de entrada e saída, mas essa relação é difícil de mapear, um desafio no qual os modelos de aprendizagem profunda têm apresentado excelente desempenho.

Os resultados obtidos mostram a adequação de modelos de aprendizagem profunda aplicados a problemas macroeconômicos. Primeiro, os modelos aprenderam a classificar os ciclos de negócios dos Estados Unidos corretamente. Em seguida, aplicando o aprendizado de transferência, eles foram obtiveram sucesso na identificação dos ciclos de negócios de dados brasileiros e europeus fora da amostra. Na mesma linha, os modelos aprenderam a estimar o hiato do produto com base nos dados americanos e obtiveram bom desempenho frente aos dados brasileiros. Em ambos os casos, a estratégia proposta surge como uma ferramenta suplementar potencial para os governos e o setor privado conduzirem suas atividades à luz das condições econômicas nacionais e internacionais. Além disso, o aprendizado profundo se mostrou adequado para mapear variáveis de baixa frequência a partir de dados de alta frequência para interpolar, distribuir e extrapolar séries temporais por séries relacionadas. A aplicação dessa técnica em dados brasileiros mostrou-se compatível com benchmarks baseados em outras técnicas.

**Palavras-chave**: Ciclo de negócios. Hiato do produto. Transferência de aprendizagem. Aprendizagem profunda. Conversão de frequência.

# ABSTRACT

Limited datasets and complex nonlinear relationships are among the challenges that may emerge when applying econometrics to macroeconomic problems. This research proposes deep learning as an approach to transfer learning in the former case and to map relationships between variables in the latter case. Several machine learning techniques are incorporated into the econometric framework, but deep learning remains focused on time-series forecasting. The approach proposed here is also related to pattern recognition, but where deep learning has achieved state-of-the-art performance: progressively using multiple layers to extract higher-level features from the raw input.

Firstly, transfer learning is proposed as an additional strategy for empirical macroeconomics. Although macroeconomists already apply transfer learning when assuming a given a priori distribution in a Bayesian context, estimating a structural VAR with signal restriction and calibrating parameters based on results observed in other models, to name a few examples, advance in a more systematic transfer learning strategy in applied macroeconomics is the innovation we are introducing. When developing economics modeling strategies, the lack of data may be an issue that transfer learning can fix. We start presenting theoretical concepts related to transfer learning and proposed a connection with a typology related to macroeconomic models. Next, we explore the proposed strategy empirically, showing that data from different but related domains, a type of transfer learning, helps identify the business cycle phases when there is no business cycle dating committee and to quick estimate an economic-based output gap. In both cases, the strategy also helps to improve the learning when data is limited. The approach integrates the idea of storing knowledge gained from one region's economic experts and applying it to other geographic areas. The first is captured with a supervised deep neural network model, and the second by applying it to another dataset, a domain adaptation procedure. Overall, there is an improvement in the classification with transfer learning compared to baseline models. To the best of our knowledge, the combined deep and transfer learning approach is underused for application to macroeconomic problems, indicating that there is plenty of room for research development.

Secondly, since deep learning methods are a way of learning representations, those that are formed by the composition of multiple non-linear transformations, to yield more abstract representations, we apply deep learning for mapping low-frequency from high-frequency variables. There are situations where we know, sometimes by construction, that there is a relationship be-tween input and output variables, but this relationship is difficult to map, a challenge in which deep learning models have shown excellent performance.

The results obtained show the suitability of deep learning models applied to macroeconomic problems. First, models learned to classify United States business cycles correctly. Then, applying transfer learning, they were able to identify the business cycles of out-of-sample Brazilian and European data. Along the same lines, the models learned to estimate the output gap based on the U.S. data and obtained good performance when faced with Brazilian data. In both cases, the proposed strategy emerges as a potential supplementary tool for governments and the private sector to conduct their activities in the light of national and international economic conditions. Additionally, deep learning proved adequate for mapping low-frequency variables from high-frequency data to interpolate, distribute, and extrapolate time series by related series. The application of this technique to Brazilian data proved to be compatible with benchmarks based on other techniques.

**Keywords**: Business cycle. Output gap. Transfer learning. Deep learning. Frequency conversion.

# LIST OF FIGURES

# LIST OF TABLES

# CONTENTS

# 1    INTRODUCTION

Econometrics, the branch of economics concerned with applying statistical and mathematical methods to economic problems, is a relevant and constantly expanding area of research. One of the questions always open is where to use a particular method, especially when presenting remarkable performance in other areas of knowledge, as deep learning nowadays.

Limited datasets and complex nonlinear relationships are among the challenges that may emerge when applying econometrics to macroeconomic problems. This research proposes deep learning as an approach to transfer learning in the former case and to map relationships between variables in the latter case. Several machine learning techniques are incorporated into the econometric framework, but deep learning remains focused on time-series forecasting. The approach proposed here is also related to pattern recognition, but where deep learning has achieved state-of-the-art performance: progressively using multiple layers to extract higher-level features from the raw input.

A major assumption in machine learning (ML) and data mining algorithms is that the training and future data must be in the same feature space and have the same distribution. However, in many real-world applications, this assumption may not hold (Pan and Yang, 2010). Can we use information from adult humans to train an intelligent system for diagnosing infant heart disease? Such a problem is known as transfer learning. The population of interest is called the target domain, for which labels are usually not available and training a classifier might be not possible. However, if data from a similar population is available, it could be used as a source of additional information (Kouw and Loog, 2018, 2). For example, we may find that learning to recognize apples might help to recognize pears. Similarly, learning to play the electronic organ may help facilitate learning the piano. The study of transfer learning (TL) is motivated by the fact that people can intelligently apply knowledge learned previously to solve new problems faster or with better solutions (Pan and Yang, 2010, 2). Thus, TL refers to the situation where what has been learned in one setting (e.g., distribution P1) is exploited to improve generalization in another setting (say, distribution P2). The learner must perform two or more different tasks, but it is assumed that many of the factors that explain the variations in P1 are relevant to the variations that need to be captured for learning P2 (Goodfellow et al., 2016, 534).

Artificial intelligence (AI) has recently gained considerable prominence due to performances in autonomous vehicles, intelligent robots, image and speech recognition, automatic translations, and medical and law usage (Makridakis, 2017). In Economics, the application of machine learning methods, an AI technique, is not new, and in a way, it has followed the phases of use in other areas. This has extended from the earliest attempts in the 1940s, followed by the rising expectations and the results in the 1960s, through the period of frustration in the 1970s, to the continuity of its use by a small group of researchers in the 1980s, and the resurgence in the 1990s (Stergiou and Siganos, 2011). Finally, from the beginning of the 21st century, significant progress has been observed in many areas, attracting attention and research funding.

Meanwhile in Macroeconomics, according to Vines and Wills (2018), during the Great Moderation (mid-1980s to 2007), the New Keynesian Dynamic Stochastic General Equilibrium (DSGE) model had become the benchmark model: the one taught to students at the start of the first-year graduate macro course. But the benchmark model has limitations. What new ideas are needed (Vines and Wills, 2018, 2)?

In Chapter 2, we present the typology of macroeconomic models introduced by Blanchard (2018) and the transfer learning setting proposed by Pan and Yang (2010). If this is true that the transfer knowledge capacity, or more broadly the representation learning, observed in other scientific areas must hold in economics, we face an underutilized tool that can contribute to applied macroeconomics. Although macroeconomists already apply transfer learning when assuming a given a priori distribution in a Bayesian context, estimating a structural VAR with signal restriction and calibrating parameters based on results observed in other models, to name a few examples, advance in a more systematic transfer learning strategy in applied macroeconomics is the new idea we are introducing.

We proceed, in Chapter 3, with the empirical setup and a literature review on the algorithms used for transfer learning, and we present in more detail our choice, deep neural networks, or deep learning, because it has shown excellent performance for transfer learning, but also because it is significantly less black box than it was in the past. Since feature importance is relevant in economics, neural networks usually are not the first choice. However, interpretability in AI is an active area of research with significant advances, both by researchers' determination and by the requirements of companies and governments to adopt these models for decision-making. Partial Dependence Plots, Permutation Feature Importance, Shapley Value, and Integrated Gradients are available methods to compute feature relevance, to name a few. Then, in Chapters 4 and 5, we advance empirically by exploring transfer learning capability when applied to business cycle identification and output gap estimation.

Lastly, in Chapter 6 we explore deep learning for mapping the low-frequency gross domestic product from high-frequency variables because there are situations where we know, sometimes by construction, that there is a relationship between input and output variables, but this relationship is di cult to map, a challenge in which deep learning models have shown excellent performance, especially towards a data-centric view, where we hold the code fixed and interactively improve the data.

## 2       TRANSFER LEARNING AND MACROECONOMICS

### 2.1       Macroeconomic models

More than forty years ago, when Sims (1980) proposed vector autoregressions (VARs) as an alternative strategy for empirical macroeconomics[1], he asserted that existing econometric analysis strategies related to macroeconomics were subject to a number of serious objections, some recently formulated, some old. For instance, empirical macroeconomists sometimes express frustration at the limited amount of information in economic time series, and it does not infrequently turn out that models reflecting rather different behavioral hypotheses fit the data about equally well (Sims, 1980, 15).

Recently, Vines and Wills (2018), in the Rebuilding Macroeconomic Theory Project, claims that the need to change macroeconomic theory is similar to the situation in the 1930s, at the time of the Great Depression, and in the 1970s, when in inflationary pressures were unsustainable. They asked several leading macroeconomists to describe how the benchmark New Keynesian model might be rebuilt in the wake of the 2008 crisis (Vines et al., 2018).

As one can see, it has been an area in constant evolution, but still with many challenges. The number of macroeconomic models is countless, and we see no restriction to use, or at least to try, the strategy proposed in Section 2.2 to any of them. Still, here we restrict our environment to the typology proposed by Blanchard (2018). According to him (Blanchard, 2018, 43-44), there should be five kinds of general equilibrium models: a common core, plus foundational theory, policy, toy, and forecasting models. The different classes of models have a lot to learn from each other, but the goal of full integration has proven counterproductive. We need different macroeconomic models for different purposes. One type is not better than the others. (Blanchard, 2018, 52-53) explains each type:

---

1   VARs are useful statistical devices for evaluating alternative macroeconomic models. His suggestion has stood the test of time well. In the early days, VARs played an important role in the evaluation of alternative models. They continue to play that role today (Christiano, 2012).

1. The purpose of the DSGE models is to explore the macro implications of distortions or sets of distortions. To allow for a productive discussion, they must be built around a largely agreed-upon common core. Each model then explores additional distortions, be it bounded rationality, asymmetric information, different forms of heterogeneity, etc. These models should aim to be close to reality, but not through ad hoc additions and repairs, such as arbitrary and undocumented higher-order costs introduced only to deliver more realistic lag structures. Fitting reality closely should be left to policy models.

2. Foundational models' purpose is to make a deep theoretical point, likely of relevance to nearly any macro model, but not pretending to capture reality closely. Some examples of this type are the consumption-loan model of Paul Samuelson, the overlapping generation model of Peter Diamond, the models of money by Neil Wallace or Randy Wright, the equity premium model of Edward Prescott and Rajnish Mehra, the search models of Peter Diamond, Dale Mortensen, and Chris Pissarides.

3. The purpose of policy models is to help policy, study the dynamic effects of specific shocks, and explore alternative policies. For these models, capturing actual dynamics is clearly essential. So is having enough theoretical structure that the model can be used to trace the effects of shocks and policies. But the theoretical structure must by necessity be looser than for DSGE: aggregation and heterogeneity lead to much more complex aggregate dynamics than a tight theoretical model can hope to capture. Old-fashioned policy models started from theory as motivation and then let the data speak, equation by equation. Some new-fashioned models start from a DSGE structure and then let the data determine the richer dynamics. In any case, for this class of models, the rules of the game here must be different than for DSGEs. Does the model t well, for example, in the sense of being consistent with the dynamics of a VAR characterization? Does it capture well the effects of past policies? Does it allow us to think about alternative policies?

4. Examples of toy models are the many variations of the IS-LM model, the Mandell-Fleming model, the RBC model, and the New Keynesian model. As the list indicates, some may be only loosely based on theory, others more explicitly so. But they have the same purpose.

They allow for a quick first pass at some questions and present the essence of the answer from a more complicated model or a class of models. For the researcher, they may come before writing a more elaborate model or after, once the elaborate model has been worked out. How close they remain formally to theory is not a relevant criterion here. They are art as much science. But art is of much value.

5. The purpose of forecasting models is straightforward: give the best forecasts. And this is the only criterion by which to judge them. If a theory is useful in improving the forecasts, then the theory should be used. If it is not, it should be ignored. The issues are then statistical, from over-parameterization to how to deal with the instability of the underlying relations.

## 2.2    An additional strategy for empirical macroeconomics

When developing one or more modeling strategies described in Section 2.1, the lack of data may be an issue that transfer learning helps to fix. As we show empirically in chapters 4 and 5, data from different but related domains, a type of transfer learning, help improve the task at hand.

Although macroeconomists already apply transfer learning when assuming a given a priori distribution in a Bayesian context, estimating a structural VAR with signal restriction and calibrating parameters based on results observed in other models, to name a few examples, advance in a more systematic transfer learning strategy in applied macroeconomics is our proposition. Then, for empirical macroeconomics, it should be feasible to apply transfer learning on data as an additional strategy to either (Pan and Yang, 2010, 5):

- instance-transfer: re-weight some labeled data in the source domain for use in the target domain.

- feature-representation-transfer: find a good feature representation that reduces difference between the source and the target domains and the error of classification and regression models.

- parameter-transfer: discover shared parameters or priors between the source domain and target domain models, which can bene t for transfer learning.

- relational-knowledge-transfer: build mapping of relational knowledge between the source and the target domains.

Similar to the characterization of toy models by Blanchard (2018), applying transfer learning to macroeconomics is an art as much science, but art is of much value. The first step in developing such an approach is collecting, organizing, and evaluating the task-related available data. As a data-driven method, transfer learning performance relies on the quality of the data. Next, the researcher must choose one or more machine learning algorithms that will learn with the data. Pan and Yang (2010) and Weiss et al. (2016) review some options, and in Section 3.1.1, we describe deep learning, our choice. Finally, one evaluates the results, transfers learning, and compares it with the outcomes of other modeling strategies if it is the case.

In the conventional transfer learning approach, we first train a base network on a base dataset and task, and then we repurpose the learned features or transfer them to a second target network to be trained on a target dataset and task. This process will tend to work if the features are general, meaning suitable to both base and target tasks, instead of specific to the base task (Yosinski et al., 2014). Figure 1 illustrates these dynamics[2].

The foregoing discussion is based on a relevant, comprehensive survey about transfer learning by Pan and Yang (2010). Consider that a *domain* $\mathcal{D}$ consist of two components: a feature space $\mathcal{X}$ and a marginal probability distribution $P(X)$, where $X = \{x_1, \ldots, x_n\} \in \mathcal{X}$. For example, if our learning task is a document classification, and each term is taken as a binary feature, then $\mathcal{X}$ is the space of all term vectors, $x_i$ is the $i^{th}$ term vector corresponding to some documents, and $X$ is a particular learning sample. In general, if two domains are different, then they may have different feature spaces or different marginal probability distributions. Given a specific domain, $\mathcal{D} = \{\mathcal{X}, P(X)\}$, a *task* consists of two components: a label space $\mathcal{Y}$ and an objective predictive function $f(\cdot)$, denoted by $\mathcal{T} = \{\mathcal{Y}, f(\cdot)\}$, which is not observed but can be learned from the training data, which consist of pairs $\{x_i, y_i\}$, where $x_i \in X$ and $y_i \in \mathcal{Y}$. The function $f(\cdot)$ can be used to predict the corresponding label, $f(x)$, of a new instance $x$. From a probabilistic viewpoint, $f(x)$ can be written as $P(y|x)$. In our document classification example, $\mathcal{Y}$ is the set of all labels, which is True, False for a binary classification task, and $y_i$ is "True" or "False". For simplicity, let's consider the case where there is one source domain $\mathcal{D}_S$, and one target domain, $\mathcal{D}_T$, as this is by far the most popular of the research works in the literature. More specifically, lets denote the *source domain data* as $\mathcal{D}_S = \{(x_{S_1}, y_{S_1}), \ldots, (x_{S_{n_S}}, y_{S_{n_S}})\}$, where $x_{S_i} \in \mathcal{X}_S$ is the data instance and $y_{S_i} \in \mathcal{Y}_S$ is the corresponding class label. In the document classification example, $\mathcal{D}_S$ can be a set of term vectors together with their associated true or false class labels. Similarly, lets denote the *target domain data* as $\mathcal{D}_T = \{(x_{T_1}, y_{T_1}), \ldots, (x_{T_{n_T}}, y_{T_{n_T}})\}$, where the input $x_{T_i}$ is in $\mathcal{X}_T$ and $y_{T_i} \in \mathcal{Y}_T$ is the corresponding output. In most cases, $0 \leq n_T \ll n_S$ (Pan and Yang, 2010, 3).

---

[2] The base networks (top two rows) are trained using standard deep learning procedures on datasets A and B. The labeled rectangles (e.g., WA1) represent the weight vector learned for that layer, with the color indicating which dataset the layer was originally trained on. The vertical, ellipsoidal bars between weight vectors represent the activations of the network at each layer. The target networks (bottom two rows) represent transfer learning strategies. The first n weight layers of the network (in this example, n = 3) are copied from a network trained on one dataset (e.g., A), and then the entire network is trained on the other dataset (e.g., B). Usually, the first n layers are either locked during training or allowed to learn.

Figure 1 - Transfer learning overview - reproduced from Yosinski et al. (2014).

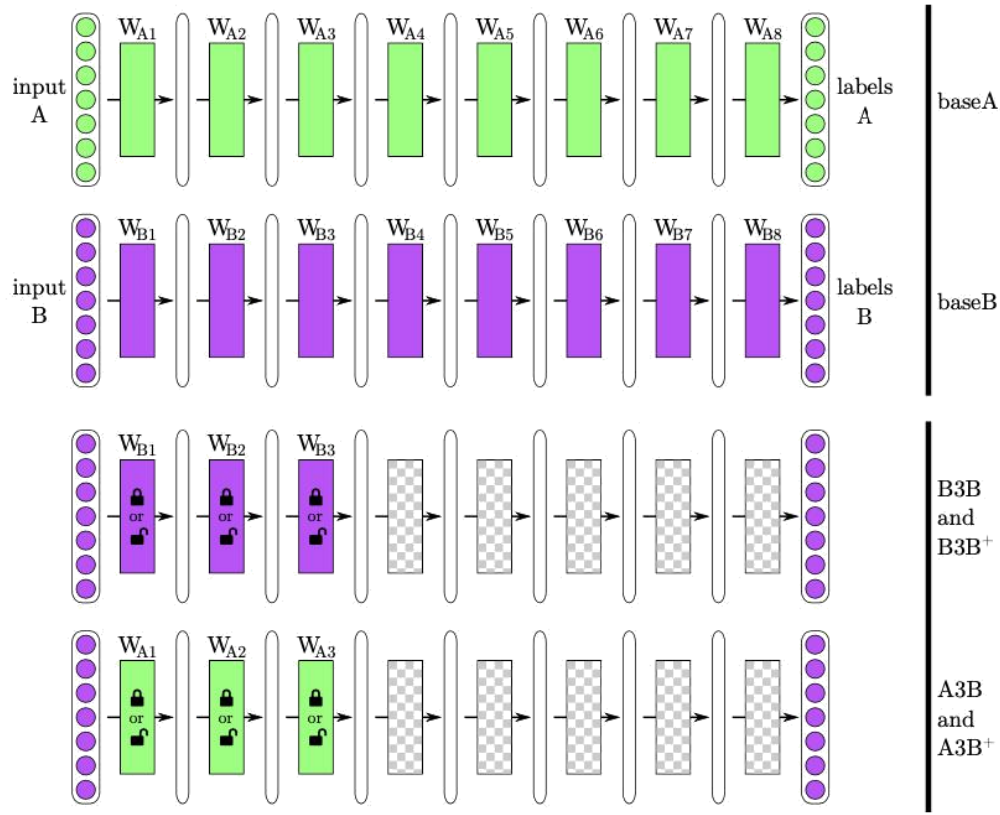

**Definition 1** *(Transfer Learning)* Given a source domain $\mathcal{D}_S$ and learning task $\mathcal{T}_S$, a target domain $\mathcal{D}_T$ and learning task $\mathcal{T}_T$, *transfer learning* aims to help improve the learning of the target predictive function $f_T(\cdot)$ in $\mathcal{D}_T$ using the knowledge in $\mathcal{D}_S$ and $\mathcal{T}_S$, where $\mathcal{D}_S \neq \mathcal{D}_T$, or $\mathcal{T}_S \neq \mathcal{T}_T$.

In the above definition, a domain is a pair $\mathcal{D} = \{\mathcal{X}, P(X)\}$. Thus, the condition $\mathcal{D}_S \neq \mathcal{D}_T$ implies that either $\mathcal{X}_S \neq \mathcal{X}_T$ or $P_S(X) \neq P_T(X)$. Similarly, a task is defined as a pair $\mathcal{T} = \{\mathcal{Y}, P(Y|X)\}$. Thus, the condition $\mathcal{T}_S \neq \mathcal{T}_T$ implies that either $\mathcal{Y}_S \neq \mathcal{Y}_T$ or $P(Y_S|X_S) \neq P(Y_T|X_T)$. When the target and source domains are the same, i.e., $\mathcal{D}_S = \mathcal{D}_T$, and their learning tasks are the same, i.e., $\mathcal{T}_S = \mathcal{T}_T$, the learning problem becomes a traditional machine learning problem. When the domains are different, then either (1) the feature spaces between the domains are different, i.e., $\mathcal{X}_S \neq \mathcal{X}_T$, or (2) the feature spaces between the domains are the same but the marginal probability distributions between domain data are different, i.e., $P(X_S) \neq P(X_T)$, where $X_{S_i} \in \mathcal{X}_S$ and $X_{T_i} \in \mathcal{X}_T$. Pan and Yang (2010) summarized the relationship between traditional machine learning and various transfer learning settings, categorizing transfer learning under three sub-settings (Table 2.1), then we can read **Definition 1** as: Given a source domain $\mathcal{D}_S$ and learning task $\mathcal{T}_S$, a target domain $\mathcal{D}_T$ and learning task $\mathcal{T}_T$,

1. *inductive transfer learning* aims to help improve the learning of the target predictive function $f_T(\cdot)$ in $\mathcal{D}_T$ using the knowledge in $\mathcal{D}_S$ and $\mathcal{T}_S$, where $\mathcal{T}_S \neq \mathcal{T}_T$.

2. *transductive transfer learning* aims to help improve the learning of the target predictive function $f_T(\cdot)$ in $\mathcal{D}_T$ using the knowledge in $\mathcal{D}_S$ and $\mathcal{T}_S$, where $\mathcal{D}_S \neq \mathcal{D}_T$ and $\mathcal{T}_S = \mathcal{T}_T$.

3. *unsupervised transfer learning* aims to help improve the learning of the target predictive function $f_T(\cdot)$ in $\mathcal{D}_T$ using the knowledge in $\mathcal{D}_S$ and $\mathcal{T}_S$, where $\mathcal{D}_S \neq \mathcal{D}_T$ and $\mathcal{Y}_S$ and $\mathcal{Y}_T$ are not observable.

Table 1 - Relationship between Traditional Machine Learning and Transfer Learning

| Learning Settings | Source and Target Domains | Source and Target Tasks |
|---|---|---|
| Traditional Machine Learning | the same | the same |
| Inductive Transfer Learning | the same | different but related |
| Transductive Transfer Learning | different but related | the same |
| Unsupervised Transfer Learning | different but related | different but related |

## 2.2.1 Inductive Transfer Learning

**Definition 2** (*Inductive Transfer Learning*) Given a source domain $\mathcal{D}_S$ and a learning task $\mathcal{T}_S$, a target domain $\mathcal{D}_T$ and a learning task $\mathcal{T}_T$, *inductive transfer learning* aims to help improve the learning of the target predictive function $f_T(\cdot)$ in $\mathcal{D}_T$ using the knowledge in $\mathcal{D}_S$ and $\mathcal{T}_S$, where $\mathcal{T}_S \neq \mathcal{T}_T$.

Based on the above definition, a few labeled data in the target domain are required as the training data to *induce* the target predictive function. This setting has two cases: (1) labeled data in the source domain are available; (2) labeled data in the source domain are unavailable while unlabeled data in the source domain are available. Most transfer learning approaches in this setting focus on the former case (Pan and Yang, 2010, 4-5). Inductive transfer learning can be used to instance-transfer, feature-representation-transfer, parameter-transfer and relational-knowledge-transfer.

### 2.2.2 Transductive Transfer Learning

**Definition 3** *(Transductive Transfer Learning)* Given a source domain $\mathcal{D}_S$ and a corresponding learning task $\mathcal{T}_S$, a target domain $\mathcal{D}_T$ and a corresponding learning task $\mathcal{T}_T$, *transductive transfer learning* aims to improve the learning of the target predictive function $f_T(\cdot)$ in $\mathcal{D}_T$ using the knowledge in $\mathcal{D}_S$ and $\mathcal{T}_S$, where $\mathcal{D}_S \neq \mathcal{D}_T$ and $\mathcal{T}_S = \mathcal{T}_T$.

In addition, some unlabeled target domain data must be available at training time. This definition covers the work of Arnold et al. (2007), since they considered *domain adaptation*, where the difference lies between the marginal probability distribution of source and target data, i.e., the tasks are the same, but the domains are different (Pan and Yang, 2010, 8). Transductive transfer learning can be used to instance-transfer and feature-representation-transfer, and it is the setting we explore in chapters 4 and 5 in two macroeconomic problems well suited for this approach, business cycle identification and output gap estimation, respectively. The intuitive behind the feature-representation-transfer is to learn a good feature representation for the target domain. In this case, the knowledge used to transfer across domains is encoded into the learned feature representation. With the new feature representation, the performance of the target task is expected to improve significantly.

### 2.2.3 Unsupervised Transfer Learning

**Definition 4** *(Unsupervised Transfer Learning)* Given a source domain $\mathcal{D}_S$ with a learning task $\mathcal{T}_S$, a target domain $\mathcal{D}_T$ and a corresponding learning task $\mathcal{T}_T$, *unsupervised transfer learning* aims to help improve the learning of the target predictive function $f_T(\cdot)$ in $\mathcal{D}_T$ using the knowledge in $\mathcal{D}_S$ and $\mathcal{T}_S$, where $\mathcal{T}_S \neq \mathcal{T}_T$ and $\mathcal{Y}_S$ and $\mathcal{Y}_T$ are not observable.

Based on this definition, no labeled data are observed in the source and target domains in training, since in unsupervised transfer learning, the predicted labels are latent variables, such as clusters or reduced dimensions. Unsupervised transfer learning can be used to feature-representation-transfer.

### 2.2.4    Transfer Learning settings and Macroeconomic typologies

As mentioned in 2.2.2, we apply transductive transfer learning in chapters 4 and 5. However, depending on the problem type and the available data, the path may di er, as shown in Figure 2.

Figure 2 - Different settings of transfer - reproduced from Pan and Yang (2010)

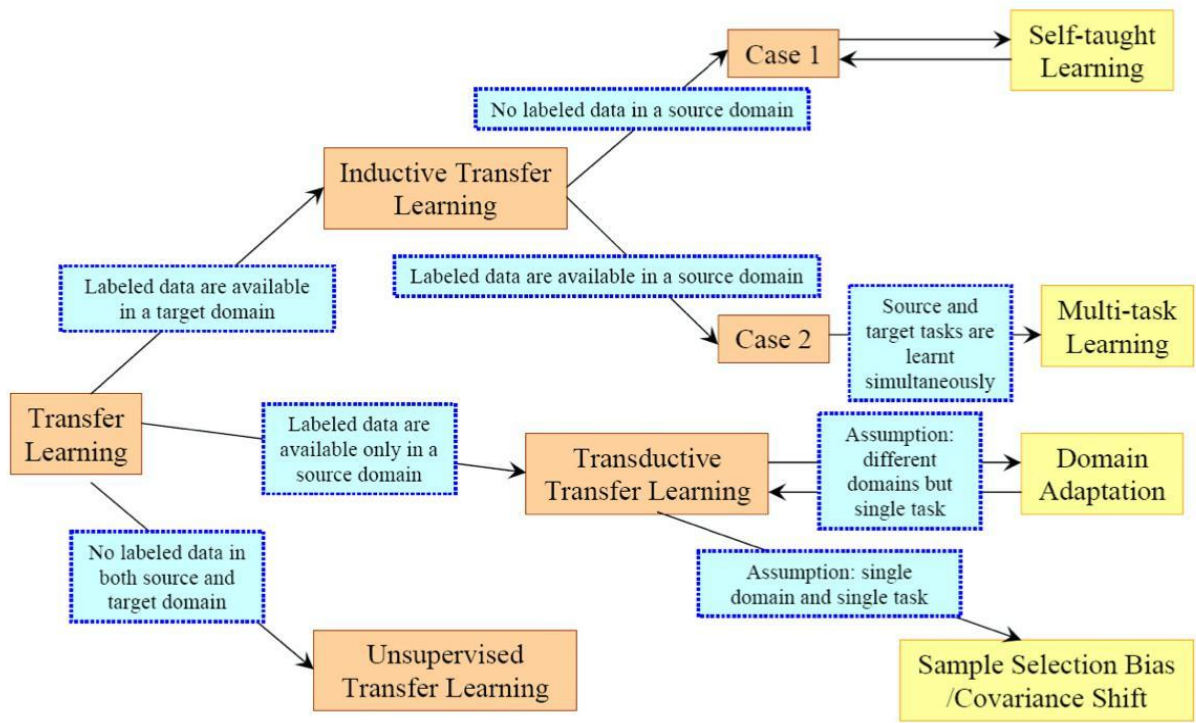

Starting through the shortest path, when there is no labeled data in both source and target domain, generally a situation in which we have data, possibly even a lot of data, but not a clear definition or classification of the existing relationships. This scenario is one of the most challenging and one of the most promising, as it is minimal or even no human intervention in assigning the relationship between the data. Some possible macroeconomics applications are clustering for regional studies and dimensionality reduction for input variables.

Next, Figure 2 shows two related cases to inductive transfer learning, where labeled data are available in a target domain, and source and target domains are the same. In case 1, the self-taught learning, there is no labeled data in a source domain, while in case 2, the multi-task learning, labeled data are available in a source domain, where the source and target tasks are learned simultaneously. From a macroeconomics perspective, relevant application could be finding priors, or initialization parameters for models, or applying natural language processing (NLP) for sentiment analysis on monetary policy committee statements. Suppose a sentiment analysis problem in which the researcher wants to classify monetary policy committee statements of a given country in hawkish or dovish. Still, there is insufficient textual data because it is a recently implemented regime in the country. The researcher can then use a model previously trained with data from other countries and then use an inductive transfer learning strategy.

Lastly, the two situations of transductive transfer learning, where labeled data are available only in a source domain with two alternative assumptions: different domains but single task, the domain adaptation case, and single domain and task, the sample selection bias/covariance shift. The latter is a recurring problem in macroeconomics and, therefore, with well-developed frameworks, which does not prevent considering transfer learning as an additional option. The former case, domain adaptation, is what we explore in chapters 4 and 5. Not all combinations between the macroeconomic models (Blanchard, 2018) typologies and the transfer learning (Pan and Yang, 2010) settings will necessarily succeed, but certainly, each one can be explored. In Figure 3 we speculate about some possibilities.

Figure 3 - Transfer Learning settings and Macroeconomic typologies

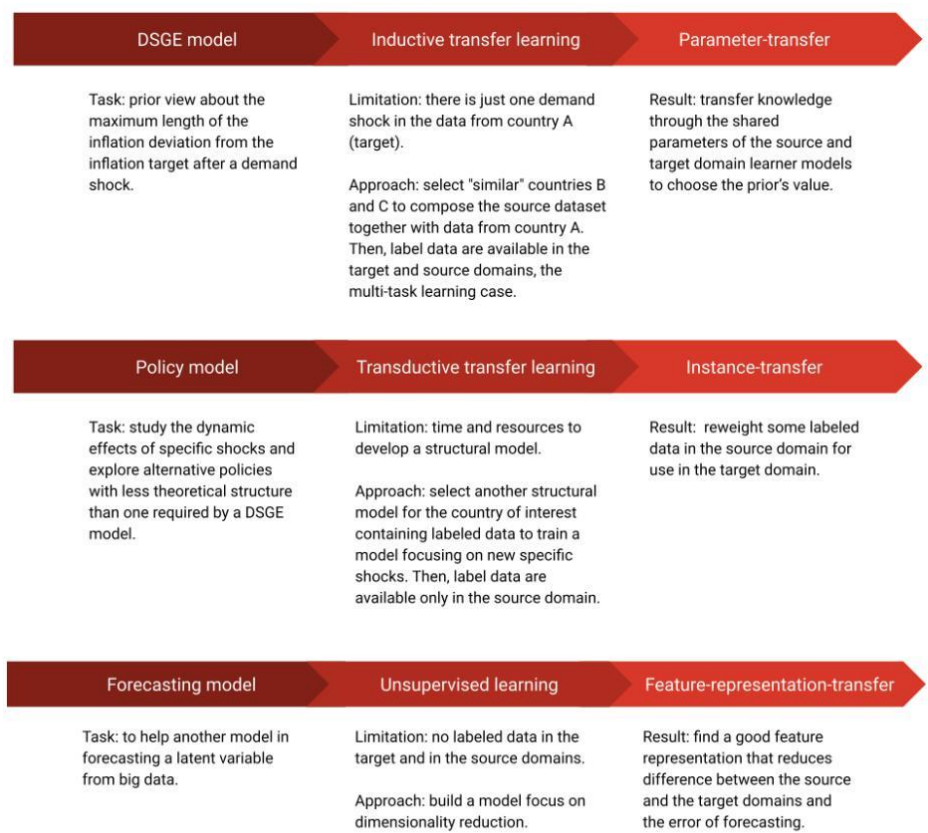

| DSGE model | Inductive transfer learning | Parameter-transfer |
|---|---|---|
| Task: prior view about the maximum length of the inflation deviation from the inflation target after a demand shock. | Limitation: there is just one demand shock in the data from country A (target).<br><br>Approach: select "similar" countries B and C to compose the source dataset together with data from country A. Then, label data are available in the target and source domains, the multi-task learning case. | Result: transfer knowledge through the shared parameters of the source and target domain learner models to choose the prior's value. |

| Policy model | Transductive transfer learning | Instance-transfer |
|---|---|---|
| Task: study the dynamic effects of specific shocks and explore alternative policies with less theoretical structure than one required by a DSGE model. | Limitation: time and resources to develop a structural model.<br><br>Approach: select another structural model for the country of interest containing labeled data to train a model focusing on new specific shocks. Then, label data are available only in the source domain. | Result: reweight some labeled data in the source domain for use in the target domain. |

| Forecasting model | Unsupervised learning | Feature-representation-transfer |
|---|---|---|
| Task: to help another model in forecasting a latent variable from big data. | Limitation: no labeled data in the target and in the source domains.<br><br>Approach: build a model focus on dimensionality reduction. | Result: find a good feature representation that reduces difference between the source and the target domains and the error of forecasting. |

# 3 EMPIRICAL SETUP

## 3.1 Algorithms for Transfer Learning

After discovering which knowledge can be transferred, learning algorithms need to be developed to transfer the knowledge (Pan and Yang, 2010, 3). To name a few, Dai et al. (2007a) proposed a boosting algorithm, TrAdaBoost, which is an extension of the AdaBoost algorithm, to address the inductive transfer learning problems, Jebara (2004) proposed to select features for multi-task learning with Support Vector Machines (SVM), Ruckert and Kramer (2008) designed a kernel-based approach to inductive transfer, which aims at finding a suitable kernel for the target data, Mihalkova et al. (2007) proposed an algorithm that transfers relational knowledge with Markov Logic Networks (MLN) across relational domains, Dai et al. (2007b) extended a traditional Naive Bayesian classifier for the transductive transfer learning problems. See Pan and Yang (2010) and Weiss et al. (2016) for many other examples.

Due to the availability of various machine learning (ML) methods and considering that we are, especially in economics, in the explanatory era of its applications, works often apply several ML approaches to a specific dataset to compare their performances, a strategy known as horse-race, as in Tiffin (2016), Cook and Hall (2017), Garcia et al. (2017), Gu et al. (2018), and Piger (2020). Makridakis et al. (2018) go further to compare various non-ML and ML forecasting methods. Another usual approach, our choice, is selecting an appropriate strategy for the specific task in advance. Deep learning is the most suitable choice for having achieved state-of-the-art performance in pattern recognition and because, according to Bengio (2012), deep learning seems well suited to transfer learning because it focuses on learning representations and, in particular, on abstract representations which ideally disentangle the factors of variation present in the input. Learning representations of the data make it easier to extract useful information when building classifiers or other predictors, and, in the case of probabilistic models, a good representation is often one that captures the posterior distribution of the underlying explanatory factors for the observed input (Bengio et al., 2013). Additionally, deep learning is significantly less black box than it was in the past. Since feature importance is relevant in economics, neural networks usually are not the first choice. However, interpretability in AI is an active area of research with significant advances, both by researchers' determination and by the requirements of companies and governments to adopt these models for decision-making. Partial Dependence Plots, Permutation Feature Importance, Shapley Value, and Integrated Gradients are available methods to compute feature relevance, to name a few. We did not advance in feature relevance estimation here, but it is already possible when necessary.

Table 2 - Out-of-Sample Evaluation Metrics for Alternative Classifiers

| Classifier | QPS | AUROC |
|---|---|---|
| **Naïve Bayes** | | |
| Narrow | 0.058 | 0.990 |
| Real Activity | 0.064 | 0.974 |
| Broad | 0.074 | 0.968 |
| **kNN** | | |
| Narrow | 0.030 | 0.989 |
| Real Activity | 0.033 | 0.978 |
| Broad | 0.055 | 0.990 |
| **Random Forest / Extra Trees** | | |
| Narrow | 0.034 | 0.988 |
| Real Activity | 0.032 | 0.988 |
| Broad | 0.036 | 0.989 |
| **Boosting** | | |
| Narrow | 0.043 | 0.980 |
| Real Activity | 0.037 | 0.978 |
| Broad | 0.039 | 0.982 |
| **LVQ** | | |
| Narrow | 0.043 | 0.938 |
| Real Activity | 0.046 | 0.930 |
| Broad | 0.038 | 0.952 |
| **DFMS** | | |
| Narrow | 0.041 | 0.997 |
| Real Activity | 0.047 | 0.992 |
| **Ensemble** | | |
| Narrow | 0.034 | 0.992 |
| Real Activity | 0.029 | 0.993 |
| Broad | 0.031 | 0.993 |

Reproduced from Piger (2020).

Our choice of a specific algorithm can also be understood in the context of the data-centric versus model-centric view. According to Ng (2021), in a model-centric view, we hold the fixed data and interactively improve code/model: take the data you have and develop a model that does as well as possible on it. On the other hand, in the data-centric view, we hold the code fixed and interactively improve the data: the quality of data is paramount and, with tools to improve the data quality, multiple models can have a good performance. To cite an example, Piger (2020) compared the performance of various machine learning techniques for business cycle prediction in the United States, finding similar results among them (Table 2), indicating that those excellent results were obtained mainly due to the quality of the data used to ingest into the models.

### 3.1.1  Deep Learning

A deep neural network, also known as deep learning (DL), is an artificial neural network (ANN) with multiple layers hidden between the input and output layers (Bengio, 2009). Deep learning is a sub-field within machine learning that is based on algorithms for learning multiple levels of representation in order to model complex relationships among data (Deng and Yu, 2014, 200). In essence, almost all DL algorithms can be described as a combined specification of a data set, a cost function, an optimization[3] procedure, and a model (Goodfellow et al., 2016). The analytical function corresponding to one of the simplest forms of an ANN, the feed-forward network, can be written as follows (Bishop, 1994, 118-9). In a feed-forward network having two layers there are $d$ inputs, $M$ hidden units and $c$ output units, where $x$ and $y$ are the network input and output variables, respectively. The output of the $j$th hidden unit is obtained by first forming a weighted linear combination of the $d$ input values, and adding a bias, to give

$$a_j = \sum_{i=1}^{d} w_{ji}^{(1)} x_i + w_{j0}^{(1)} \; . \tag{3.1}$$

---

[3] Deep learning algorithms involve optimization in many contexts. See (Goodfellow et al., 2016, 271-325) for a comprehensive discussion.

Here $w_{ji}^{(1)}$ denotes a weight in the first layer, going *from* input *i to* hidden unit *j*, and $w_{j0}^{(1)}$ denotes the bias for hidden unit *j*. The bias term for the hidden units is made explicit by the inclusion of an extra input variable $x_0$ whose value is permanently set at $x_0 = 1$. This can be represented analytically by rewriting (3.1) in the form

$$a_j = \sum_{i=0}^{d} w_{ji}^{(1)} x_i \; .$$

(3.2)

The activation of hidden unit *j* is then obtained by transforming the linear sum in (3.2) using an activation function $g(\cdot)$ to give

$$z_j = g\!\left(a_j\right) \; .$$

(3.3)

The outputs of the network are obtained by transforming the activations of the hidden units using a second layer of processing elements. Thus, for each output unit *k*, we construct a linear combination of the outputs of the hidden units of the form

$$a_k = \sum_{j=1}^{M} w_{kj}^{(2)} z_j + w_{k0}^{(2)} \; .$$

(3.4)

Again, we can absorb the bias into the weights to give

$$a_k = \sum_{j=0}^{M} w_{kj}^{(2)} z_j \; ,$$

(3.5)

which can be represented by including an extra hidden unit with activation $z_0 = 1$ . The activation of the *k*th output unit is then obtained by transforming this linear combination using a non-linear activation function, to give

$$y_k = \tilde{g}(a_k) \; .$$

(3.6)

Here we have used the notation $\tilde{g}(\cdot)$ for the activation function of the output units to emphasize that this need not be the same function as used for the hidden units. If we combine (3.2), (3.3), (3.5) and (3.6) we obtain an explicit expression for the complete function in the form

$$y_k = \tilde{g}\left(\sum_{j=0}^{M} w_{kj}^{(2)} \, g\left(\sum_{i=0}^{d} w_{ji}^{(1)} x_i\right)\right).$$

(3.7)

These models are called feed-forward because information flows through the function being evaluated from inputs $x$, through the intermediate computations used to define the function, and finally to the output target $y$. There are no feedback connections in which the outputs of the models are fed back into itself. When processing sequential data is required, there are alternatives like recurrent neural networks (RNN), that process an input sequence one element at a time, maintaining in their hidden units a *state vector* that implicitly contains information about the history of all past elements of the sequence (LeCun et al., 2015). The long short-term memory (LSTM), a gated RNN, is one of the most effective sequence models used in practical applications (Goodfellow et al., 2016, 404). The LSTM forward propagation equations for a shallow recurrent network architecture are given below (Goodfellow et al., 2016, 405-406). Instead of a unit that simply applies an element-wise nonlinearity to the affine transformation of inputs and recurrent units, LSTM recurrent networks have "LSTM cells" that have an internal recurrence (a self-loop), in addition to the outer recurrence of the RNN. Each cell has the same inputs and outputs as an ordinary recurrent network, but also has more parameters and a system of gating units that controls the flow of information. The most important component is the state unit $s_i^{(t)}$, which has a linear self-loop. The self-loop weight, or the associated time constant, is controlled by a **forget gate** unit $f_i^{(t)}$, for time step $t$ and cell $i$, which sets this weight to a value between 0 and 1 via sigmoid unit:

$$f_i^{(t)} = \sigma\left(b_i^f + \sum_j U_{i,j}^f x_j^{(t)} + \sum_j W_{i,j}^f h_j^{(t-1)}\right),$$

(3.8)

where $x^{(t)}$ is the current input vector and $h^{(t)}$ is the current hidden layer vector, containing the outputs of all the LSTM cells, and $b^f$, $U^f$, $W^f$ are respectively biases, input weights, and recurrent weights for the forget gates. The LSTM cell internal state is thus updated as follows, but with a conditional self-loop weight $f_i^{(t)}$:

$$s_i^{(t)} = f_i^{(t)} s_i^{(t-1)} + g_i^{(t)} \sigma \left( b_i + \sum_j U_{i,j} x_j^{(t)} + \sum_j W_{i,j} h_j^{(t-1)} \right), \qquad (3.9)$$

where $b$, $U$ and $W$ respectively denote the biases, input weights, and recurrent weights into the LSTM cell. The **external input gate** unit $g_i^{(t)}$ is computed similarly to the forget gate, with a sigmoid unit to obtain a gating value between 0 and 1, but with its own parameters:

$$g_i^{(t)} = \sigma \left( b_i^g + \sum_j U_{i,j}^g x_j^{(t)} + \sum_j W_{i,j}^g h_j^{(t-1)} \right). \qquad (3.10)$$

The output $h_i^{(t)}$ of the LSTM cell can also be shut off, via the **output gate** $q_i^{(t)}$, which also uses a sigmoid unit for gating:

$$h_i^{(t)} = tanh\left( s_i^{(t)} \right) q_i^{(t)} , \qquad (3.11)$$

$$q_i^{(t)} = \sigma \left( b_i^o + \sum_j U_{i,j}^o x_j^{(t)} + \sum_j W_{i,j}^o h_j^{(t-1)} \right), \qquad (3.12)$$

which has parameters $b^o$, $U^o$ and $W^o$ for its biases, input weights and recurrent weights, respectively.

### 3.1.2     Cost function

According to (Goodfellow et al., 2016, 173-4), the cost functions for neural networks are more or less the same as those for other parametric models, such as linear models. The total cost function used to train a neural network will often combine cost functions with a regularization term in order to prevent overfitting[4]. Hence, we can think of three situations concerning the overfitting problem, where the model family being trained either 1) exclude the correct data-generating process corresponding to underfitting and inducing bias, or 2) match the true data generating process, or 3) include the generating process but also many other possible generating processes - the overfitting regime where variance rather than bias dominates the estimation error. The goal of regularization is to take a model from the third regime into the second regime (Goodfellow et al., 2016, 224). Regularization is any modification we make to a learning algorithm that is intended to reduce its generalization error but not its training error. Regularization is one of the central concerns of the field of machine learning, rivaled in its importance only by optimization (Goodfellow et al., 2016, 117).

### 3.1.3     Optimization procedure

When we use a machine learning algorithm, we sample the training set, and then use it to choose the parameters to reduce training set error. We then sample the test set. Ander this process, the expected test error is greater than or equal to the expected value of training error. The factors that determine how well a machine learning algorithm will perform are its ability to: 1) make the training error small; 2) make the gap between training and test error small (Goodfellow et al., 2016, 109).

Back-propagation is a method to calculate a gradient that is needed in the calculation of the weights to be used when training the network. Back-propagation is a special case of an older and more general technique called automatic differentiation. In the context of learning, back-propagation is commonly used by the optimization algorithm to adjust the weight of neurons by calculating the gradient of the loss function. This technique is also sometimes called backward propagation of errors, because the error is calculated at the output and distributed back through the network layers. (Goodfellow et al., 2016, 213) describe the general back-propagation procedure.

---

[4] See Goodfellow et al. (2016), Chapters 6 and 7 for further discussion.

The problem of determining the capacity of a deep learning model is especially difficult because the capabilities of the optimization algorithm limit the effective capacity, and we have little theoretical understanding of the general non-convex optimization problems involved in deep learning (Goodfellow et al., 2016, 112). Even though most practitioners, for many years, believed that local minima were a common problem plaguing neural network optimization, today that does not appear to be the case. The problem remains an active area of research, but experts now suspect that, for sufficiently large neural networks, most local minima have a low-cost function value, and that it is not essential to find the correct global minimum rather than to find a point in parameter space that has low but not minimal cost (Goodfellow et al., 2016, 282).

## 3.2    Practical Methodology

Since machine learning practitioners cannot prove, at least until now, which model is the best for each situation, we search for successful deep artificial neural networks that can exhibit small differences between training and test performance (Zhang et al., 2016). According to Goodfellow et al. (2016), successful applying deep learning techniques requires more than just a good knowledge of what algorithms exist and the principles that explain how they work. A good machine learning practitioner also needs to know how to choose an algorithm for a particular application and how to monitor and respond to feedback obtained from experiments to improve a machine learning system. See (Goodfellow et al., 2016, 416-435) for more details about performance metrics, default baseline models, determining whether to gather more data, selecting hyperparameters, and debugging strategies.

The models in Chapters 4, 5 and 6 were built using TensorFlow[5], an interface for expressing machine learning algorithms, and an implementation for executing such algorithms. TensorFlow is flexible and can be used to express a wide variety of algorithms, including training and inference algorithms for deep neural network models. It has been used to conduct research and deploy machine learning systems into production across more than a dozen areas of computer science and other elds (Abadi et al., 2015).

---

[5] https://www.tensorflow.org/.

### 3.2.1    Hyperparameters space

The primary architectural considerations are choosing the depth of the network and the width of each layer. Deeper networks are often able to use far fewer units per layer and far fewer parameters, as well as frequently generalizing to the test set, but they also tend to be harder to optimize. The ideal network architecture for a task must be found via experimentation guided by monitoring the validation set error (Goodfellow et al., 2016, 194). The optimal set of hyperparameters was obtained using Hyperband (Li et al., 2018) from Keras Tuner[6]. We delimited the grid search by hyperparameters according to Table 3 by looking for parsimonious models and observing what is generally adopted in the literature. As a robustness test, we retrained the models with others loss functions, the squared hinge (Chapter 4) and the adaptive gradient algorithm (AdaGrad) (Duchi et al., 2011) (Chapter 5), finding similar results.

Table 3 - Hyperparameters space

| Hyperparameter | Grid search | Step |
|---|---|---|
| Dense network depth | 1 - 4 | +1 |
| Dense network hidden units | 16 - 256 | 2 |
| LSTM network hidden units | 16 - 256 | 2 |
| Lambda regularizer | 0.0001 - 0.01 | 10 |
| Learning rate | 0.0001 - 0.01 | 10 |
| Activation function | ReLU, Tanh, Sigmoid | OR |

---

[6] https://www.tensorflow.org/tutorials/keras/keras_tuner.

## 3.3    Feature representation

A good feature representation should be able to reduce the difference in distributions between domains as much as possible, while at the same time preserving essential properties of the original data (Pan et al., 2011). When we transfer learning, though, there are additional issues. The challenge is to overcome the differences between the domains so that a classifier trained on the source domain generalizes well to the target domain, but generalizing across distributions can be difficult, and it is not clear which conditions have to be satisfied for a classifier to perform well (Kouw and Loog, 2018). From a macroeconometrics perspective, we could assume that the input variables' marginal contributions in explaining the business cycle phases are similar. However, they are probably not equal for the different economic areas, as well as the respective input variables' sample covariance matrix. Since we are applying transductive transfer learning (subsection 2.2.2), we can presume two circumstances (Pan and Yang, 2010, 4): (1) the feature spaces between the source and target domains are different; (2) the feature spaces between domains are the same, but the marginal probability distributions of the input data are different. The latter case is related to domain adaptation. Our approach handles this well when the models successfully build a common representation space for the different domains, which can be empirically verified when there are labels for the target domain, allowing evaluation by some error criteria. Otherwise, it remains an open question.

### 3.3.1   Robustness: baseline and benchmark models

Our baseline models are a logistic model (Chapter 4), a linear regression model (Chapter 5), and deep learning models without the transfer learning approach. They function as a reference to verify if negative transfer (Torrey and Shavlik, 2009) is not occurring, a situation in which the transfer reduces the learning. And, as benchmarks, we compare our results with those released by entities that apply widely accepted techniques to identify the business cycle and estimate the output gap, notably the Centre for Economic Policy Research (CEPR), the Brazilian Business Cycle Dating Committee (CODACE), and the Central Bank of Brazil. For Chapter 6, as bench-marks, we have the Chow and Lin (1971) method, the GDP Monitor from Instituto Brasileiro de Economia (2015), and the Central Bank Economic Activity Index (IBC-Br). The choice of these benchmarks is relevant because of the risk of overfitting[7]. In other words, no matter how complex the relationships are between the data, there is always the risk of a neural network mapping the entire data set, resulting in overfitting, and subsequently failing in terms of generalization by not correctly classifying data not previously observed. That is why the test set is so relevant. That is why our strategy of using data from a country other than the one used for network training as a test set is a critical factor in evaluating the models' performance.

### 3.3.2    Real-time data

Depending on the task at hand, the real-time data approach assumes relevance when building the database. For instance, if the objective is forecasting the business cycle, as in Piger (2020), we could use vintages containing the first release of data related to economic activity, whose posterior revision can significantly affect the results. Notably, the models proposed here do not require adjustments to deal with real-time data. Instead, use real-time data to ingest the models, which will scan for the optimal results based on this information because machine learning models today largely reflect the patterns of their training data (Google, 2021).

### 3.3.3    Economics school of thought

A supervised deep learning model learns from the features-targets without the need for strong assumptions about its relation, which prevents the expert from choosing an underlying economics school of thought to set up a model, although the variable selection might represent it. The algorithm maps the input-output relation variables according to the training and validation data. For instance, when we choose the National Bureau Economic Research (NBER)[8] turning points classification data as output label (Chapter 4), or the output gap based on the potential output estimated by the U.S. Congressional Budget Office (CBO)[9] as the target (Chapter 5), the deep learning algorithm will learn from them how to classify the business cycle and to measure the output gap, implicitly following the same schools of thought.

---

[7] According to , for every $k \geq 2$, there exists a neural network with ReLU activations of depth $k$, width $O(n/k)$, and $O(n + d)$ weights that can represent any function on a sample of size $n$ in $d$ dimensions.

[8] The most followed classification for U.S. business cycle.

[9] An economic-based estimation. See Shackleton (2018).

### 3.4 Small data, big data, good data

A deep learning model is capable of handling a large number of explanatory variables (features) as indicator series (Chapter 6). However, there are some caveats, as the more features we use during learning, the more data preprocessing and computer processing e orts will be required. For example, if we train for the United States output gap estimation using dozens of features, we will need the same quantity of features to transfer learning to other datasets or additional preprocessing strategies, as in Jackson and Rege (2019) that have fed an ANN with dynamic factors. Although we are not working with big data, the models proposed here can deal with high-dimensional structured data. For unstructured data, however, modifications to the models would be necessary. However, it is essential to emphasize that good data is sought, whether small or big.

It is also true when transfer learning, that improves learning in a new task by transferring knowledge from a related task that has already been learned (Torrey and Shavlik, 2009). More specifically, domain adaptation can be considered a special set of transfer learning that aims at transferring shared knowledge across different but related tasks or domains. An example of the current state of the art is transfer learning with MobileNets (Howard et al., 2017). The MobileNet model is an open-source code that learned how to classify 1,000 categories from more than 1,500,000 images. It has some verification points to start from freezing the previous points' parameters and training a new model that recognizes images that are not necessarily among the 1,000 categories previously learned. This new model would start from the representation learning present in the parameters transferred from MobileNet. For example, in a simple exercise, presenting only 100 new images of gestures used for the game stone, paper, and scissors, a transfer learning model can identify each new category with high precision[10]. Without transfer learning from MobileNet, the number of images needed to train this new model would be much higher.

#### 3.4.1 Robustness: locked and unlocked models

When transfer learning, the weight layers of the network for Euro and Brazilian data (Chapter 4), or just Brazilian data (Chapter 5), were copied from the network trained on the U.S.

---

[10] https://github.com/lmoroney/dlaicourse/tree/master/TensorFlow, in TensorFlow Deployment, Course 1, Week 4.

data, as in the last row of Figure 1, except that we do not retrain the parameters. It is the locked models and works as if these two datasets function as out-of-sample, also known as label extension. Additionally, we unlock the last layers and retrain the parameters, a way of relaxing the macroeconometrics assumptions about input variables as mentioned before. In a big data context, as in the MobileNet example mentioned before, unlocked models are expected to perform better than locked models because they use information from the target dataset for training. However, in smaller databases, like ours, negative transfer learning can occur, a situation in which some information from the target dataset ends up introducing noise that damages the performance on training step.

## 3.5      Cross-sectional data into feed-forward neural network

As in the dynamic factor model with Markov-switching introduced by Chauvet (1998), the deep learning approach accounts for the idea of business cycles as the simultaneous, asymmetrical, and nonlinear movement of economic activity in various sectors. This data-driven framework is flexible enough to be training with different features, the independent variables, and targets, the dependent variables. The option for a feed-forward network as a deep learning model, which represents memory-less models, derives from the focus on contemporary movement between the selected input variables and the business cycle and the output gap to explore the informational content of their cross-section distributions. This implies disregarding the time dependence observed on the variables and shuffling the data to break it. The resulting model accounts just for coincident relations. One advantage is the reduced computational cost for training compared to configurations that map the temporal dependency, which may not be very significant in small datasets but is very relevant when working with big data. Although we are not working with big data, we are looking for network configurations that are suitable for this case. Another relevant point is that we emphasize the focus on contemporary relationships between inputs and labels, as our main objective is to transfer learning (Chapters 4 and 5) and convert the frequency of variables (Chapter 6). In this sense, we reinforce the need to understand the limitations of our approach, emphasizing that models trained to detect correlations should not be used to make causal inferences or imply that they can.

3.5.1    Robustness: sequential approach with LSTM

To account for time dependency as an alternative strategy, we create additional models in Chapters 4 and 5 including a long short-term memory (LSTM) layer.

## 3.6    Normalization, unbalanced, seasonality, stationarity, missing values

A few more remarks about the empirical setup. Normalization, or standardization, is an important step that has empirically shown positive results for learning in deep learning models. In Chapter 6, for instance, the scale of the features we used differs by several orders of magnitude, making learning slower or even less optimal if we did not adopt normalization when pre-processing the data. In the same sense, in classification models such as in Chapter 4, it must be considered when the database is unbalanced, that is, disproportional between positive and negative events, making adjustments in the base or in the model, such as setting an initial bias.

Seasonality and stationarity are issues to be considered. Although the diagnosis itself and how to correct these characteristics is not always trivial, the better the quality of the data ingested in the model, the better its performance should be. In our case, we chose to run alternative models in Chapter 4 because input variables show an increasing trend over time (Appendix A), although this is not always feasible, especially when dealing with big data. Thus, we use seasonally adjusted, level, and first difference data. In the case of FNN models with a cross-sectional approach, an additional step in data pre-processing is shuffling the data, which also minimizes seasonality and non-stationarity effects. In Chapter 6, where there is a mix of stationary and non-stationary, seasonal, and non-seasonal variables, we run models with observed data and, alternatively, with year over year transformation.

Finally, missing values can be handled prior to ingesting data into the models, with traditional strategies to impute data, or during model training, more common in cases of unstructured data such as texts, where the model itself can predict a missing word in a sentence. Naturally, the late procedure could increase the time and affect the performance during the model training. In the current state of the art, treating missing values in the pre-processing data phase with the most suitable technique for each circumstance is the most recommended approach.

# 4        TRANSFER LEARNING FOR BUSINESS CYCLE

Monitoring business cycle phases is a traditional task in applied macroeconomics. Progressive market integration has induced a worldwide interest in analyzing cyclical fluctuations through economic indicators Chauvet (2001). Changes in exchange rates, outputs, consumption, inflation, and interest rates in different parts of the world can influence the effectiveness of government policies and the competitive position of businesses, even those not directly related to international operations (Chauvet and Yu, 2006, p.43). As a result, a wide range of techniques has been developed since the seminal work by Burns and Mitchell (1946). Recently, new approaches have emerged due to the progress in machine learning (ML) research, centering on building models that achieve better forecasting performance than the non-ML models or that identify turning points more timely (Piger, 2020).

This Chapter[11] is organized as follows. Section 4.1 presents a literature review. In Section 4.2, we discuss the implementation details of the proposed new strategy (Section 2.2) when applied to Business Cycle identification, and, in Section 4.3, the empirical findings are presented.

## 4.1      Related Literature

Business cycles are recurrent sequences of alternating phases of expansion and contraction among many economic activities (Burns and Mitchell, 1946). According to Harding and Pagan (2005), there are three ways in the literature to describe what we mean by a cycle, depending on whether the focus is on the fluctuation of the level of economic activity, the level of economic activity less a permanent component, or the growth rate of economic activity. Stock and Watson (2014) examine two approaches to identifying turning points, the (i) average-then-date, which describes the dating of reference cycles using a single highly aggregated series, such as GDP, and the (ii) date-than-average, the approach of the pioneers of business cycle dating, who considered a large number of disaggregated series. Their contributions provide a nonparametric definition of a turning point and produce standard errors for the date-then-average chronologies.

---

[11] A modified version of this Chapter was published in Central Bank of Brazil Working Paper Series (Chauvet and Guimaraes, 2021).

In the United States, the NBER Business Cycle Dating Committee provides a chronology of business cycle expansion and recession dates. According to Piger (2020), because the NBER methodology is not explicitly formalized, literature has worked to develop and evaluate formal statistical methods for establishing the historical dates of economic recessions and expansions in both the U.S. and international data. Estrella and Mishkin (1998), Estrella et al. (2000), Issler and Vahid (2006), Kauppi and Saikkonen (2008), Rudebusch and Williams (2009) and Fossati (2016) use an available historical indicator of the class, such as the NBER dates, to estimate the parameters of models such as logit or probit ones. This strategy is called a supervised classifier in the statistical classification literature, in contrast to unsupervised classifiers, which endogenously determine the classes. Unsupervised classifiers have also been used, with the primary example being the Markov-switching (MS) framework of Hamilton (1989), which become a relevant tool for applied work in economics. Chauvet (1998) proposes a dynamic factor model with Markov-switching (DFMS) to identify expansion and recession phases from a group of coincident indicators and Chauvet and Hamilton (2005), Chauvet and Piger (2008) and Camacho et al. (2018) evaluate the performance of variants of this DFMS model to identify NBER turning points in real-time. See Piger (2020) for a comprehensive review.

Applied ML papers related to business cycles can be separated depending on whether the main focus is predicting or identifying turning points and phases. For example, Hoptro et al. (1991), Qi (2001), Klinkenberg (2003), Nasr et al. (2007), Berge (2013), Ma (2015), Garbellano (2016), Nyman and Ormerod (2017), and James et al. (2019) have applied machine learning techniques such as artificial neural networks, support vector machines, boosting, k-nearest neighbor, and random forest to forecasting turning points, recessions, or business cycles phases mainly in the US, but also other countries[12]. These studies have generally reported some improvements over non-ML strategies. The other set of papers is concerned about identifying the turning points for real-time classification. Morik and Ruping (2002), Giusto and Piger (2017), Soybilgen (2018), Ra not and Benoit (2019) and Jackson and Rege (2019) have applied inductive logic programming, learning vector quantization, random forest, boosting, k-nearest neighbor and artificial neural networks fed with dynamic factors. Piger (2020), in a comprehensive analysis, compares five ML techniques with DFMS. These studies have reported quickly and accurately turning points identification.

---

[12] United Kingdom, Japan, West Germany, and Lebanon.

Lastly, some literature is dedicated to the study of business cycles worldwide, as in Chauvet and Yu (2006), Cuba-Borda et al. (2018), Abberger et al. (2020), and the reference turning points of the OECD Composite Leading Indicators[13].

## 4.2    Implementation

The feature selection comprises the coincident variables indicated by NBER[14] as the fundamental: gross domestic product (GDP), income, employment, industrial production, and wholesale-retail sales. Quarterly data is adopted because this is the frequency at which some relevant variables for the classification of the business cycle are available, and at this frequency, opposed to higher ones, the data usually carry less noise, which may facilitate the training and the transfer learning. We computed the first difference of the logarithm of the input features, capturing the growth rate (Harding and Pagan, 2005, 152-154). Alternatively, we run the model without this transformation, i.e., features in level. In both cases, the features are normalized. To have a common starting point for each dataset, we restrict the series' start to eliminate missing values. For example, we do not acquire data for the Euro area before 2005 because we do not have employment data before this year for all the selected countries. Even if we used some strategy to extend this variable, as mentioned in 3.6, given that the others started in 1995, it would not be enough to incorporate another recession period according to CEPR data. We adopted a U.S. dataset for deep learning and two datasets, with data from Brazil and Europe, for transfer learning. The target values are the business cycle chronology provided by the NBER, the CODACE, and the CEPR Euro Area Business Cycle dating committees, respectively. Appendix

---

[13] Available at https://www.oecd.org/sdd/leading-indicators.

[14] The NBER does not define a recession in terms of two consecutive quarters of decline in real GDP. Rather, a recession is a significant decline in economic activity spread across the economy, lasting more than a few months, normally visible in real GDP, real income, employment, industrial production, and wholesale-retail sales. Source: https://www.nber.org/cycles.html.

A summarizes the information about all series, mostly from the Federal Reserve Economic Data (FRED)[15] dataset, provided by the Federal Reserve Bank of St. Louis. Our data les are available at https://github.com/rrsguim/PhD_Economics.

Following Piger (2020), the area under the ROC curve (AUC) is the objective function to maximize in the validation step when training the deep learning model. This metric is desirable here for being scale-invariant, measuring how well predictions are ranked, rather than their absolute values, and classification-threshold-invariant, measuring the quality of the model's predictions irrespective of what classification threshold is chosen.

Beginning with the deep learning step, we split the U.S. dataset into train, validation, and test sets. Then, we de ne a function that creates a neural network with hidden layers, ReLU as activation function, a dropout layer to reduce overfitting, and a sigmoid output layer that returns the probability of recession. Next, we retrain the model with the optimal hyperparameters, selected with Hyperband, to evaluate the results in both datasets, source and target, with binary cross-entropy as a loss function and Adam (Kingma and Ba, 2017) for optimization.

All codes used are available at https://github.com/rrsguim/PhD_Economics. In Appendix B we reproduce selected parts of the best performance transfer learning model for business cycle identification.

## 4.3    Results

Before presenting the results, it is relevant to assert that contributing to the turning points forecast improvement is not the objective of this chapter. The predicting of the next business cycle phase turning point, which becomes relevant at times like the current one when many countries are in the recession phase due to the COVID-19 pandemic, is a well-consolidated part of the business cycle research field (Section 4.1) and whose state of the art is similar to macroeconomics forecasting in general. However, when focusing on identifying the business cycle phase, we address challenges as the absence of a business cycle dating committee and the limited datasets. Another aspect to emphasize is that applying transfer learning to the Brazilian and the Euro area data is a choice related to the availability of committees that explicitly adopt the same, or very similar, classification approach than the committee (NBER) selected when training the models. This choice allows empirically evaluate the proposed method performance from labels in the source and target datasets. Good performance increases confidence to apply it to unlabeled datasets.

---

[15] https://fred.stlouisfed.org/.

The method proposed in Subsection 2.2.1 involves two stages: the first one consists of a machine learning model that learns to perform a specific task based on a dataset; the second focuses on applying that learning to another dataset. The deep learning models, our choice as mentioned in 3.1.1, have learned how to classify business cycle phases. The transfer learning performance, stage 2, proved to be superior to our expectations. Figures 4, 5, 6, and 7 consolidate the results found. Results are slightly different each time the models are run due to different compositions in the selected data sets for training, validation, and test, especially in the cross-sectional approach.

Figure 4 - Deep Learning (U.S)

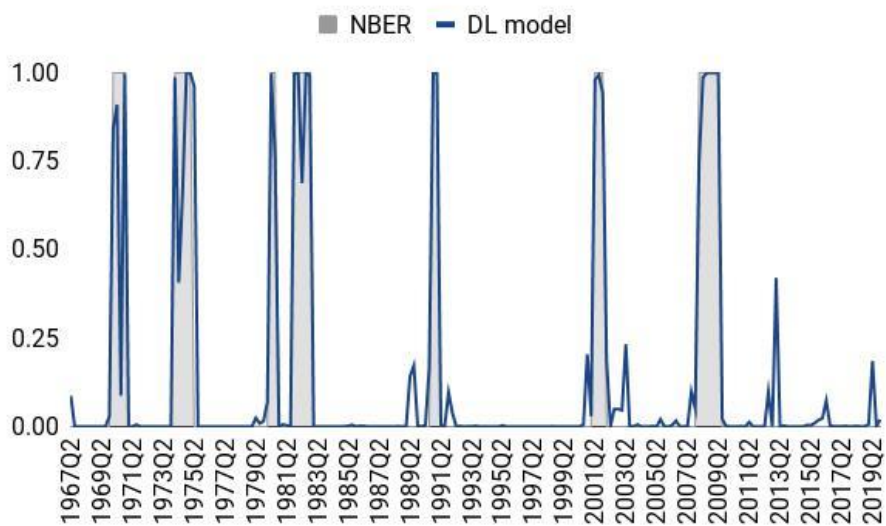

The comparison in Figure 4 shows the performance of the deep learning step: the U.S. business cycles, where shadow areas indicate recession phases according to NBER, and the feed-forward neural networks (FNN) model with the cross-sectional approach estimation, where the blue lines are showing the recession probability on each quarter. The alternative model that accounts for time dependency exhibited exceptional results as well, with almost all points correctly classified. Concerning the baseline models with U.S. data, we observed a significantly superior performance, measured by the AUC with out-of-sample data, of the deep learning models with data in first difference (1df). The outcomes of the baseline models for the nineteen selected countries in Europe (EURO-19), where there is no transfer learning step, do not motivate confidence because the FNN model has a perfect classification with AUC=1, while the alternative model that includes an LSTM layer is unable to identify crises (AUC=0.5). This point highlights one of the problems that the proposed methodology seeks to solve: identifying business cycles when data is limited. Note, in Figure 5, that there are only two recessions for the EURO-19 in the period under analysis so that models trained only on these data show inconsistent results. Regarding Brazil's outcomes, whose period contains more than one recession, the baseline models show more satisfactory performance (Figure 6). We expected these results because machine learning methods are good at pattern recognition. The results align with the other machine learning strategies adopted in the literature (Section 4.1). There is no innovation here, but the confirmation that deep learning models correctly classify the business cycle phases.

However, the transfer learning step, which is not in the business cycle literature to the best of our knowledge, exceeded our expectations. Figure 5 compares EURO-19's CEPR classification, Brazil's CODACE classification, and the transfer learning models results. These results refer to the locked models, meaning the estimates for EURO-19 and Brazil operate as if they were out-of-sample because the parameters trained with U.S. data are locked in the transfer learning phase when applied to target EURO-19 and Brazil datasets. Figure 6 presents the details, and Figure 7 a summary comparison of each models' performance. Differences in the sizes of out-of-sample data sets reflect the need for adjustments according to time series, cross-sectional, locked, and unlocked strategies. Overall, there is an improvement in the classification with transfer learning compared to baseline models. Although both showed excellent results, the LSTM models performed better than the FNN models, and locked models overperform unlocked models[16]. On the other hand, the models didn't learn well with data in level, requiring transformation. As mentioned in Section 3.2, successfully applying deep learning techniques requires monitoring and responding to feedback obtained from experiments, which is data driven. Not necessarily the same configuration will emerge from different datasets, which demands finding the best hyperparameters to perform well the task at hand.

---

[16] As mentioned in 3.4.1, better performance is generally expected on unlocked models, but the opposite can occur in cases such as when dealing with small data.

Figure 5 - Transfer Learning (Euro Area and Brazil)

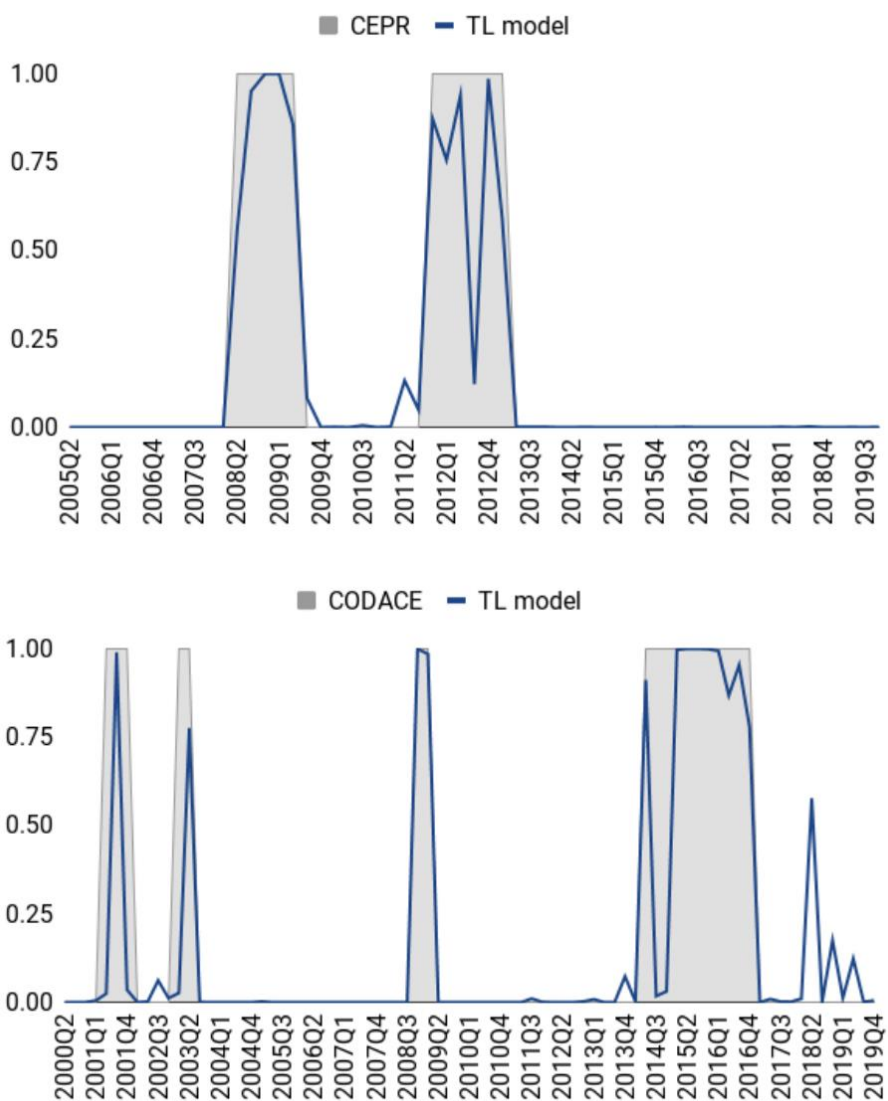

Figure 6 - Models specifications and results

| | Dropout | Hidden layers | | *Hyperband* best hyperparameters | | Target | Out-of-sample | | | | | | |
| | | LSTM | Dense | units | learning rate | | Test size | Confusion matrix | | | | | AUC |
| | | | | | | | | TN | FP | FN | TP | | |
| **Baseline models** | | | | | | | | | | | | | |
| Logit_1df_US | - | - | - | - | - | NBER | 77 | 36 | 0 | 32 | 9 | | 0.610 |
| DL_FNN_1df_US | 0.5 | 0 | 3 | 208 | 0.01 | NBER | 64 | 56 | 0 | 2 | 6 | | 0.875 |
| **DL_LSTM_1df_US** | 0.5 | 1 | 4 | 176 | 0.01 | NBER | 85 | 71 | 5 | 0 | 9 | | **0.967** |
| DL_FNN_level_US | 0.5 | 0 | 4 | 176 | 0.01 | NBER | 64 | 54 | 0 | 10 | 0 | | 0.500 |
| DL_LSTM_level_US | 0.5 | 1 | 4 | 176 | 0.01 | NBER | 85 | 76 | 0 | 9 | 0 | | 0.500 |
| DL_FNN_1df_EUR | 0.5 | 0 | 3 | 112 | 0.01 | CEPR | 18 | 15 | 0 | 0 | 3 | | 1.000 |
| DL_LSTM_1df_EUR | 0.5 | 1 | 4 | 48 | 0.0001 | CEPR | 42 | 36 | 0 | 6 | 0 | | 0.500 |
| DL_FNN_1df_BR | 0.5 | 0 | 3 | 64 | 0.01 | CODACE | 24 | 18 | 1 | 3 | 2 | | 0.674 |
| DL_LSTM_1df_BR | 0.5 | 1 | 4 | 240 | 0.001 | CODACE | 32 | 17 | 4 | 1 | 10 | | 0.859 |
| **Transfer learning models** | | | | | | | | | | | | | |
| TL_FNN_1df_EUR_locked | 0.5 | 0 | 3 | 208 | 0.01 | CEPR | 60 | 48 | 1 | 2 | 9 | | 0.899 |
| TL_FNN_1df_EUR_unlocked | 0.5 | 0 | 4 | 208 | 0.01 | CEPR | 42 | 36 | 0 | 6 | 0 | | 0.500 |
| **TL_LSTM_1df_EUR_locked** | 0.5 | 1 | 4 | 176 | 0.01 | CEPR | 60 | 48 | 1 | 1 | 10 | | **0.944** |
| TL_LSTM_1df_EUR_unlocked | 0.5 | 1 | 4 | 64 | 0.01 | CEPR | 42 | 36 | 0 | 6 | 0 | | 0.500 |
| TL_FNN_1df_BR_locked | 0.5 | 0 | 3 | 208 | 0.01 | CODACE | 80 | 60 | 2 | 6 | 12 | | 0.817 |
| TL_FNN_1df_BR_unlocked | 0.5 | 0 | 4 | 208 | 0.01 | CODACE | 24 | 16 | 3 | 1 | 4 | | 0.821 |
| **TL_LSTM_1df_BR_locked** | 0.5 | 1 | 4 | 176 | 0.01 | CODACE | 80 | 61 | 1 | 3 | 15 | | **0.909** |
| TL_LSTM_1df_BR_unlocked | 0.5 | 1 | 4 | 64 | 0.01 | CODACE | 48 | 31 | 4 | 2 | 11 | | 0.866 |

TN - true negative | FP - false positive | FN - false negative | TP - true positive | AUC - area under the ROC curve.

Figure 7 - AUC out-of-sample

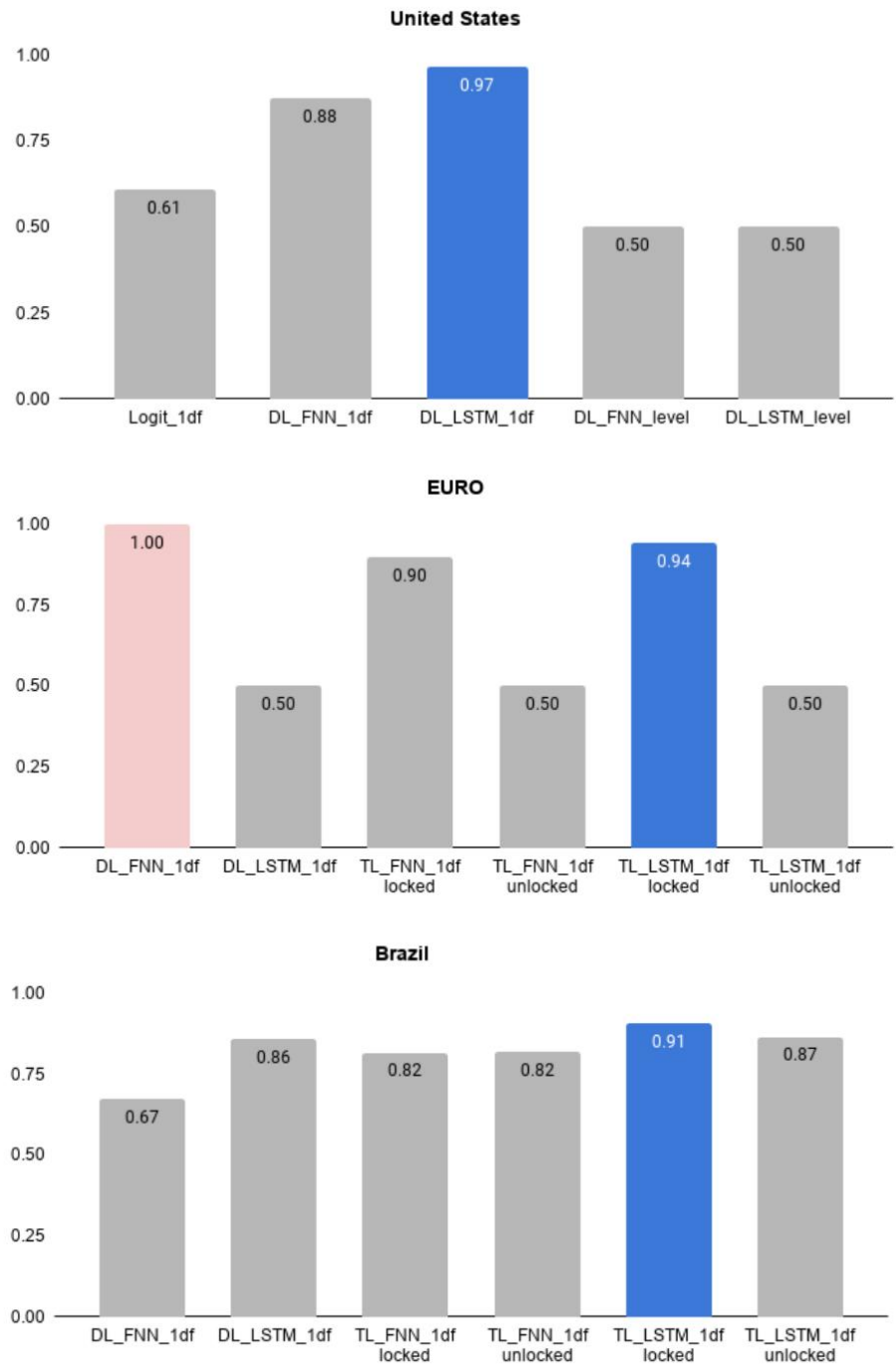

# 5      TRANSFER LEARNING FOR OUTPUT GAP

Measures of the gap between actual and potential activity are used frequently as indicators of the economic cycle and play a vital role in monetary and fiscal policy (Koske and Pain, 2008). As a result, the wide range of techniques that have been developed since the seminal work by Okun (1962) can be classified in different ways, such as between observed and unobserved components methods or between pure statistical versus economic base approaches each with its pros and cons. The emergence of new approaches applied to business cycles due to machine learning research progress (Piger, 2020) can also be explored for the output gap, given that they are related concepts.

This Chapter is organized as follows. Section 5.1 presents a literature review. In Section 5.2, we discuss the implementation details of the proposed new strategy (Section 2.2) when applied to Output Gap estimation, and, in Section 5.3, the empirical findings are presented.

## 5.1      Related Literature

How much output can the economy produce under conditions of full employment? Okun's seminal paper (Okun, 1962) begins with this question and describes two related concepts, potential output, and the output gap. Potential output differs from actual only because the potential concept depends on the assumption that aggregate demand is exactly at the level that yields a rate of unemployment equal to four percent of the civilian labor force[17]. If, in fact, aggregate demand is lower, part of potential GNP (Gross National Product, a measure of output.) is not produced; there is unrealized potential or a gap between actual and potential output (Okun, 1962). Since then, various definitions of potential output, as well as estimation strategies, have been proposed and used in the literature, depending on the investigator's objectives, although Okun's definition is still the main reference concept for economic policy-makers, including central banks (Proietti et al., 2007).

---

[17] It was a reference to full employment at that time, as he discusses in the paper.

According to Proietti et al. (2007), a useful classification is between unobserved components and observed components methods. Harvey and Jaeger (1993) trend-cycle decomposition of output and the Hodrick and Prescott (1981) HP filter are examples of univariate approaches of unobserved components, while Kuttner (1994), Gerlach and Smets (1999) and Apel and Jansson (1999) focus on multivariate approaches. Observed components methods rely on the Beveridge and Nelson (1981) decomposition and on structural vector autoregressive (VAR) models, which have been used by Blanchard and Quah (1989) and St-Amant and van Norden (1997). St-Amant and van Norden (1997) states that univariate methods, such as the HP filter, are not reliable to measure the output gap. Hybrid methods that combine univariate and structural relationships have proved to be hard to estimate, they may not be robust to reasonable alternative calibrations, and it is not easy to calculate their appropriate confidence intervals. However, methods combining estimated dynamics with structural information o er an interesting alternative. Another conceivable classification, as in D'Auria et al. (2010), is pure statistical versus economic base approaches, the latter being preferred because of the possibility of examining the underlying economic factors which are driving any observed changes in the potential output indicator and consequently the opportunity of establishing a meaningful link between policy reform measures with actual outcomes. Hamilton (2018) goes further and claims that, although the HP filter method continues today to be adopted in academic research, policy studies, and analysis by private-sector economists, it is intended to produce a stationary component from an I(4) series. However, in practice, it can fail to do so and invariably imposes a high cost. It introduces spurious dynamic relations that are purely an artifact of the filter and have no basis in the true data-generating process. He proposes a regression of the variable at date t + h on the four most recent values as an alternative approach. Okun (1962) foresaw the challenges mentioned above in his seminal work: the quantification of potential output - and output gap - is, at best, an uncertain estimate and not a rm, precise measure. Chauvet and Guimaraes (2021) proposed a transfer learning strategy to identify business cycle phases when data are limited or there is no business cycle data committee. The approach integrates the idea of storing knowledge gained from one region's economic experts and applying it to other geographic areas. Here, we adopt the same strategy to measure the output gap.

## 5.2    Implementation

The feature selection comprises variables usually incorporated in economic-based approaches: un-employment rate, capacity utilization, business cycle phase, and total factor productivity parity with U.S. when transfer learning. Quarterly data is adopted because this is the frequency at which some relevant variables for the estimation of the output gap are available, and at this frequency, opposed to higher ones, the data usually carry less noise what may facilitate the training and the transfer learning. We adopted a U.S. dataset for deep learning and data from Brazil for transfer learning. The target[18] values are calculated based on the CBO's potential output for the U.S. and the Brazilian Central Bank's output gap for Brazil. Table 7 and Figure 20 in the Appendix summarizes all series information from the FRED[19] dataset provided by the Federal Reserve Bank of St. Louis and from the Series Management System (SGS)[20] dataset provided by the Brazilian Central Bank. Our data les are available at https://github.com/rrsguim/PhD_Economics.

Beginning with the deep learning step, we split the U.S. dataset into train, validation, and test sets. Then, we de ne a function that creates a neural network with hidden layers, ReLU, tanh, or sigmoid as activation functions, and L1 (Lasso) regularization to reduce overfitting. Next, we retrain the model with the optimal hyperparameters, selected with Hyperband, to evaluate the results in both datasets, source, and target, with mean squared error as a loss function and the adaptive moment estimation (Adam) (Kingma and Ba, 2017) for optimization. All codes used and other results are available at https://github.com/rrsguim/PhD_Economics. See Appendix B for selected codes.

## 5.3    Results

Before presenting the results, it is relevant to assert that evaluate the output gaps disclosed by different sources is not the objective of this chapter. As it is an unobservable variable, there will always be some dispute about its most appropriate value. It is common for analysts to observe more than one source for decision-making. However, by focusing on learning how to estimate the output gap based on data disclosure from a selected source, we address challenges like quickly assess an economic-based output gap and the limited datasets. Another aspect to emphasize is that applying transfer learning to the Brazilian area data allows empirically evaluate the proposed method performance from labels in the source and target datasets. The theoretical model, however, enables the application to any region. Good performance increases confidence to apply it to unlabeled datasets also.

---

[18] This Chapter aims to observe the results of an empirical application to the method proposed in 2.2, and not evaluate the output gaps disclosed by different sources.

[19] https://fred.stlouisfed.org/.

[20] https://www3.bcb.gov.br/sgspub.

The method proposed in Subsection 2.2 involves two stages: the first one consists of a machine learning model that learns to perform a specific task based on a dataset; the second focuses on applying that learning to another dataset. The deep learning models, our choice as mentioned in 3.1.1, have learned how to estimate the output gap. The transfer learning performance, stage 2, proved to be superior to our expectations. Figures 8, 9, 10, and 11 consolidate the results found. Results are slightly different each time the models are run due to different compositions in the selected data sets for training, validation, and test.

Figures 8 shows the performance of the deep learning step: the U.S. output gap based on the CBO's potential output, compared with the FNN and the LSTM models. Concerning the baseline models with U.S. and Brazilian data, Figure 9, we observed a superior performance, measured by the mean absolute error (MAE) with out-of-sample data, of the feed-forward neural network (FNN) models. We expected these results because machine learning methods are good at fitting curves. There is no innovation here, but the confirmation that deep learning models correctly identify the output gap.

Figure 8 - US Output GAP - Congress Budget Office (CBO) and Deep Learning models

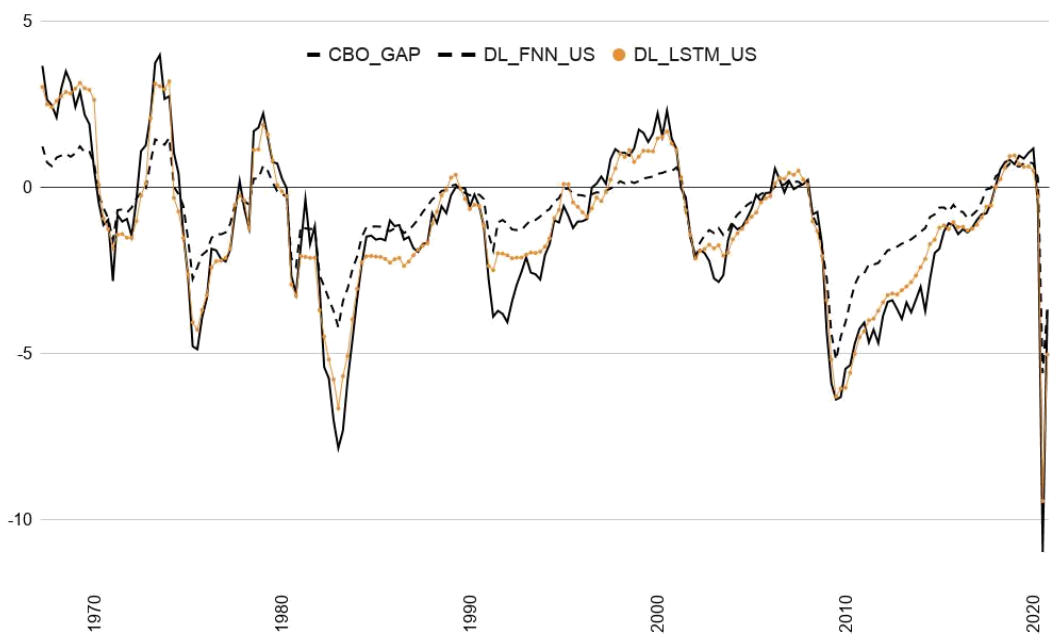

Figure 9 - Models specifications and results

| | Hidden layers | | *Hyperband* best hyperparameters | | | | | Optmizer | In-sample | | Out-of-sample | |
|---|---|---|---|---|---|---|---|---|---|---|---|---|
| | LSTM | Dense | LSTM | | Dense | | | | Size | MAE | Size | MAE |
| | | | Units | Dropout | Units | Activation | L1 lambda | Learning rate | | | | | |
| **Baseline models** | | | | | | | | | | | | |
| Linear_US_1967_2020 | - | - | - | - | - | - | - | - | Adam | 172 | 0.615 | 43 | 0.668 |
| DL_FNN_US_1967_2020 | - | 4 | - | - | 112 | ReLu | 0.001 | 0.01 | AdaGrad | 172 | 0.431 | 43 | 0.558 |
| DL_LSTM_US_1967_2020 | 1 | 4 | 128 | 0.3 | 240 | ReLu | 0.01 | 0.001 | Adam | 172 | 0.985 | 43 | 0.654 |
| DL_FNN_BR_2012-2020 | - | 4 | - | - | 112 | ReLu | 0.001 | 0.01 | AdaGrad | 27 | 0.361 | 7 | 0.366 |
| DL_LSTM_BR_2012-2020 | 1 | 4 | 16 | 0.4 | 240 | Tanh | 0.01 | 0.01 | Adam | 27 | 0.493 | 7 | 1.390 |
| **Transfer learning models for Brazil dataset 2012-2020** | | | | | | | | | | | | |
| TL_FNN_BR_locked | - | 4 | - | - | 112 | ReLu | 0.001 | 0.01 | AdaGrad | - | - | 34 | 0.646 |
| TL_FNN_BR_unlocked | - | 4 | - | - | 112 | ReLu | 0.001 | 0.01 | AdaGrad | 17 | 0.524 | 17 | 0.452 |
| TL_LSTM_BR_locked | 1 | 4 | 128 | 0.3 | 240 | ReLu | 0.01 | 0.001 | Adam | - | - | 34 | 0.614 |
| TL_LSTM_BR_unlocked | 1 | 4 | 112 | 0.4 | 224 | Tanh | 0.01 | 0.01 | Adam | 17 | 1.359 | 17 | 3.845 |

CBO - United States Congressional Budget Office | BCB - Central Bank of Brazil.

However, the transfer learning step, which is not in the output gap literature to the best of our knowledge, exceeded our expectations. Figure 10 shows a comparison between the output gap estimated by the locked transfer learning model and the one by the Central Bank of Brazil which the methodology is based on a Bayesian model estimation (Banco Central do Brasil, 2020a), a different approach. It must be emphasized that, since it refers to the locked model, Brazil's estimate operates as if it was out-of-sample because the parameters trained with U.S data are locked in the transfer learning phase, with no retraining with Brazilian data. The similarity is impressive because the machine learned, based on the U.S. data, how to estimate the output GAP and then transfer this knowledge to Brazilian data obtaining similar results of the Brazilian Central Bank specialists[21], with just one extra piece of information about Brazil, the total factor productivity (TFP) level at current purchasing power parities with the U.S. Additionally, the good performance of transfer learning is observed, despite the period under analysis showing a negative trend[22] of the output gap, indicating that the model could learn in the presence of non-stationarity. However, depending on the researcher's goals, further adjustments to the data may be necessary. Figure 11 presents a summary comparison of the performance of each model.

Figure 10 - Output GAP - Central Bank of Brazil (BCB) and Transfer Learning Model

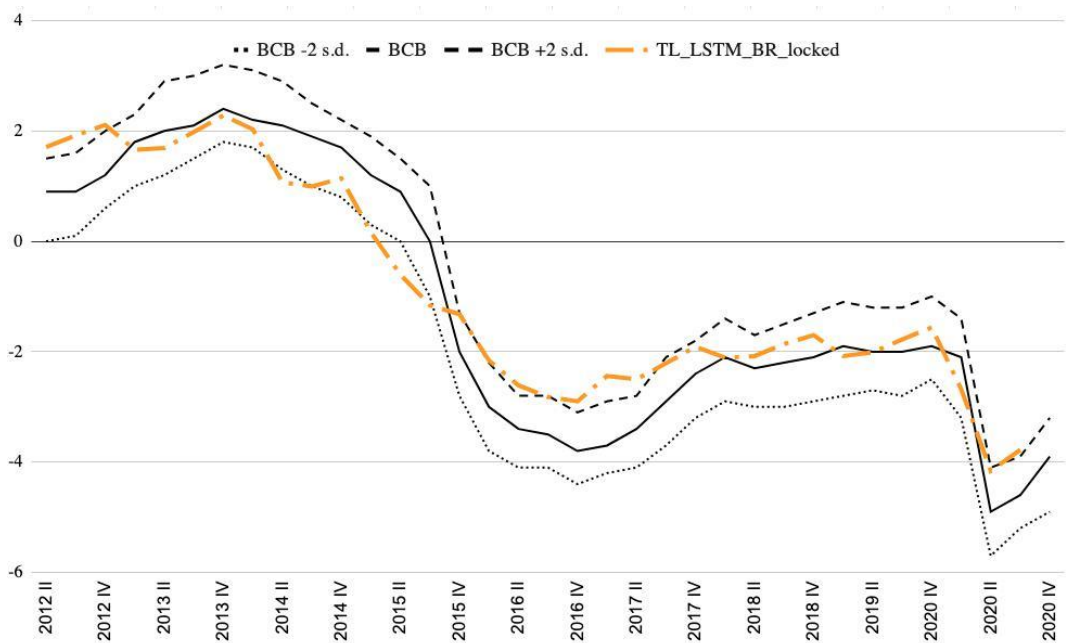

---

[21] We rebuilt the dataset based on the graph and text data at (Banco Central do Brasil, 2020b, 56) because it does not publish the output gap series.
[22] Stationarity is expected, by definition, in large series of the output gap.

Figure 11 - MAE out-of-sample

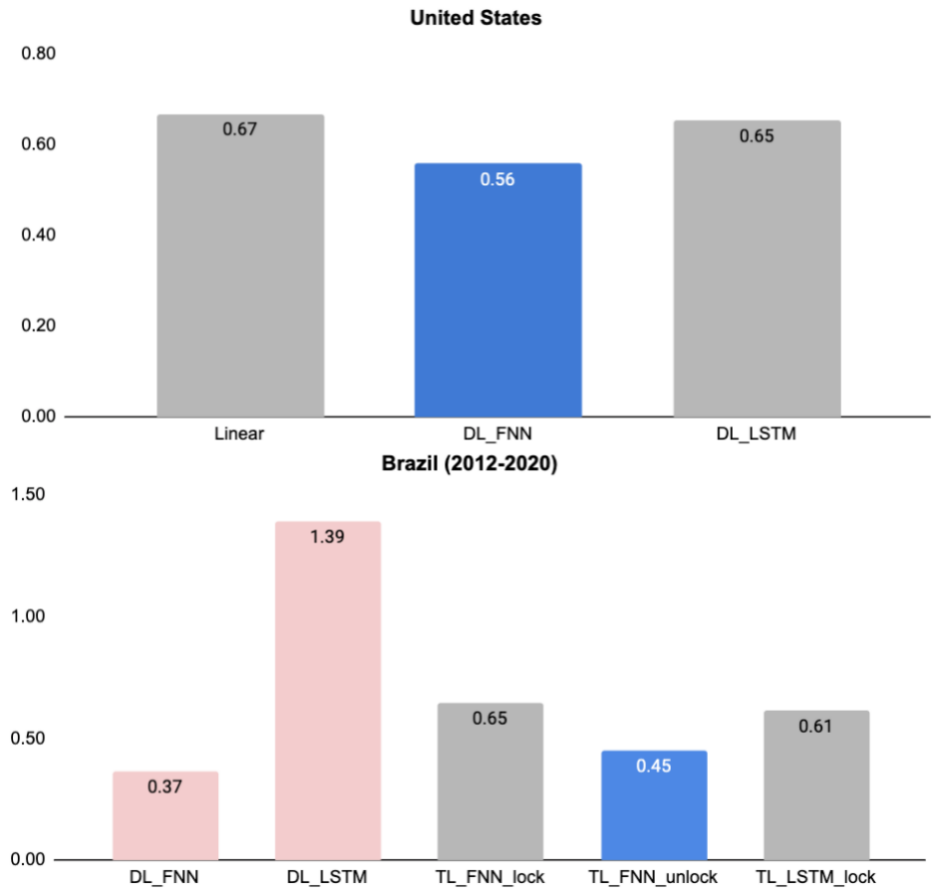

# 6 REPRESENTATION LEARNING FOR INTERPOLATION, DISTRIBUTION, AND EXTRAPOLATION OF TIME SERIES BY RELATED SERIES (RIDE)

## 6.1 Introduction

Economic and financial analysts commonly use time series modeling to predict future values, analyze their salient properties and characteristics, and monitor the current state of the economy. However, as it is not uncommon that some relevant variables are not available with the desired time and frequency, there is a research program aimed at transforming low-frequency economic variables into high-frequency variables (Pavía-Miralles, 2010). The procedure by Chow and Lin (1971), a generalization of Friedman (1962), is a seminal reference to traditional statistical methods that use related series, also called indicators. The main attraction of this approach, based on a linear regression model set of indicators with first-order stationary autoregressive errors, is its simplicity (Mazzi and Proietti, 2015). However, these methods are restricted to a limited number of indicator series due, until the recent past, to data and computation restrictions. Currently, large data sets are readily available, and models with hundreds of parameters are easily estimated (Angelini et al., 2006), allowing new approaches to open questions[23], such as using data with daily frequency, which is increasingly available[24].

The purpose of this Chapter is to apply deep learning for mapping low-frequency from high-frequency variables[25]. Since deep learning methods are a way of learning representations, those that are formed by the composition of multiple non-linear transformations, with the goal of yielding more abstract - and ultimately more useful - representations (Bengio et al., 2013), and also inspired by Chow and Lin (1971), we denominate our strategy Representation Learning for Interpolation, Distribution, and Extrapolation of Time Series by Related Series (RIDE).

---

[23] In a comprehensive analysis, Pavía-Miralles (2010) describes the main techniques used, their advantages and disadvantages, as well as the open questions related to this eld of research.

[24] (Chow and Lin, 1971, 372) mention that despite dealing only with estimating a monthly series from their quarterly data and considering related monthly series, the theory, with minimal modifications, applies to the estimation of other frequencies.

[25] We are not constructing a latent variable indicator to evaluate the state of the economy. Since the seminal article by Burns and Mitchell (1946), an extensive literature has developed along these lines, with emphasis on the dynamic factor models, e.g., Stock and Watson (2011) and Giannone et al. (2013).

## 6.2     Literature

6.2.1     Interpolation, Distribution and Extrapolation of Time Series by Related Series

Milton Friedman was an American statistician and economist awarded the Nobel Prize in Economics in 1976 and known as the father of monetarism. He was an intellectual leader with vast production. In 1962, Friedman published The Interpolation of Time Series by Related Series, a detailed analysis of the procedures adopted to construct time series related to economic phenomena. He highlighted the problem of estimating intermediate values that emerge, for example, when using data from a biennial census for annual estimation of national income or even aggregating accounting information between groups of banks that report them in different periods and frequencies. One of the most common operations performed in this process is interpolation: estimating some component for dates for which it is not directly available from known values of that component for other dates (Friedman, 1962, 729). A closely related operation, also widely used, is the distribution of a known total of a time unit among the shorter time units, of which the longest unit is composed (Friedman, 1962, 730). Another related operation is the extrapolation of some com-ponents to more recent dates to obtain current numbers before available reference data that will be used later in the interpolation (Friedman, 1962, 730). Directing his analysis to interpolation, he first states that this process, when done based only on the series being interpolated, takes the form of an old and well-known mathematical problem with extensive literature but rarely used in the construction of economics time series. The procedure most used incorporates series known or assumed to be related to the series to be interpolated (Friedman, 1962, 730). He argues, however, that this procedure has not been explored more rigorously, noting that each researcher uses their ad hoc procedure. Friedman establishes that the interpolation of a series from related series involves two steps: 1) selection of the related series to be used; 2) interpolation of the target series from related series. He clarifies that his focus is on techniques for interpolation and that he will not deal with selection (step 1). More specifically, he will deal with the restricted case where only one related series is used (Friedman, 1962, 731). After systematically presenting the methods used and associated errors and discussing various technical details involved in the choices to be made, he presents his conclusions (Friedman, 1962, 751-2), highlighting: i) mathematical interpolation and interpolation from related series are not substitute methods, they are complementary; ii) related series can improve interpolation; iii) correlation between series and deviations from trends are relevant points; iv) interpolation must be performed only on the part of the series that is unknown for the dates for which the interpolation will be performed, never generally covering the entire period.

Chow and Lin (1971) propose a unified approach to the problems of interpolation, extrapolation, and distribution that Friedman (1962), despite acknowledging to be related, dealt with them in isolation. Like Friedman (1962), they point out that the variable selection process is beyond their scope (Chow and Lin, 1971, 372). As a solution, they present the derivation of an unbiased linear best estimator, assuming that the monthly observations of the $y$ series to be estimated satisfy a multiple regression relationship with $p$ related series $x1,\ldots,x_p$. In the sample period of $3n$ months, where $n$ is the number of quarters, the relation is

$$y = X\beta + u \ , \tag{6.1}$$

where $y$ is $3n$x$1$, $X$ is $3n$x$p$ and $u$ is a random vector with zero mean and covariance matrix $V$, and

$$\beta = (X'V^{-1}X)^{-1}X'V^{-1}y \tag{6.2}$$

is the least squares estimator resulting from the regression coefficients using $n$ quarterly observations of the sample, with the vectors subscripted with a dot indicating that they refer to quarterly data, and

$$\hat{u}_. = y_. - X_.\hat{\beta} \tag{6.3}$$

is the $n$x$1$ residual vector of the regression using quarterly data.

Chow and Lin (1971) describe details of their approach, such as the treatment distinction between flow and stock series or the estimation of $V$ in three different ways. Thus, according to the authors, with a straightforward application of the unbiased linear best estimate theory to a regression model, the problems of interpolation, distribution, and extrapolation of a time series by related series can be solved in a unified way. The resulting estimator applies to all these cases. The usefulness of this method in practice, concerning the estimation of monthly economic time series, certainly depends on the validity of the assumed regression model or the possibility of finding related series that make the regression model a good approximation to reality. According to (Pavía-Miralles, 2010, 455), Chow and Lin (1971) is probably the most influential and cited article in the field of research related to interpolation, distribution, and extrapolation of time series that use related series, or indicators. This procedure has two main advantages over those that do not use indicators: i) better foundations in constructing hypotheses, and ii) they are more efficient because they adopt relevant economic and statistical information. Moreover, as the main disadvantage, it indicates that the results obtained are influenced by choice of indicators.

## 6.2.2    Artificial Neural Networks and Economics

Directing our attention to applied papers in economics, we begin with White (1988) which reports the results of a project using neural network modeling and learning techniques to search for and decode nonlinear regularities in asset price movements, based on the daily returns of the IBM common shares. In the same year, Dutta and Shekhar (1988) apply neural networks to predict the ratings of corporate bonds, finding evidence that it is a useful approach to generalization problems in such non-conservative domains. In 1991, Kuan and White (1991) published Artificial Neural Networks: An Econometric Perspective, a reference document for econometrics. According to Tkacz and Hu (1999), the use of neural networks in economics was still in its relative infancy, and the paper by Kuan and White (1991) was likely the definitive introduction of neural networks to the econometrics literature. In a business report, Schwartz (1992) states that in the world of finance, anything that provides even a slight edge over rivals can mean millions in extra profits. Thus, investment professionals are turning to gurus who o er exotic computer technologies such as neural networks and genetic algorithms. Forecasting corn futures, Kohzadi et al. (1995) found that the prediction error of a neural network model was between 18 and 40 percent lower than that of an ARIMA model, using different forecasting performance criteria. Tal and Nazareth (1995) reports that, in 1994, the Canadian Imperial Bank of Commerce replaced its index-based Leading Indicators with a neural network-based system and that the performance to that date had been very encouraging. Portugal (1995) provides an empirical comparative evaluation of the performance of the artificial neural network to the problem of economic time series forecast, performing exercises in the gross industrial output of the state of Rio Grande do Sul (Brazil) to find mixed results. Swanson and White (1997) applied neural networks to predict macroeconomic variables, contrasting different linear and nonlinear models using a wide array of out-of-sample forecasting performance indicators. Herbrich et al. (1998) provide an overview of existing economic applications of neural networks, distinguished in three types: classification of economic agents, time series prediction, and the modeling of boundedly rational agents. Tkacz and Hu (1999) pointed out that linear models are, in effect, constrained neural network models, forecast output growth using neural networks and compare the performance of such models with traditional linear specifications. They conclude that the best neural network models outperform the best linear models by between 15% and 19% of their data, implying that neural network models can be exploited for noticeable gains in forecast accuracy. According to them, the gains in forecast accuracy seem to originate from the ability of neural networks to capture asymmetric relationships. Blake (1999) makes straightforward analogies between artificial neural networks and models more familiar to economists. He also applies artificial neural network to model GDP growth using variables that could be expected to lead the growth cycle or predict likely future growth disturbances for six major European economies: France, Germany, Italy, the Netherlands, Spain and the UK.

More recently, Varian and Choi (2009) found that simple seasonal autoregressive models and fixed-effects models that include relevant Google Trends variables tend to outperform models that exclude these predictors. Analyzing data from Chile, Carriere-Swallow and Labbe (2010) presents evidence that the inclusion of information on Google search queries improves both the in- and out-of-sample accuracy of car sales models. Varian (2014) states that econometricians, statisticians, and data-mining specialists often seek insights that can be extracted from the data, and while the most common tool used for summarization is (linear) regression analysis, machine learning offers a set of tools that can usefully summarize various sorts of nonlinear relationships in the data. To overcome the problem of reliable data on economic livelihoods in the developing world, Jean et al. (2016) developed a method of estimating consumption expenditure and asset wealth using high-resolution satellite imagery. With a similar goal to ours but applying other techniques of machine-learning named elastic net and random forest, Tiffin (2016) built a nowcasting indicator for Lebanon's GDP and achieved good results within an ensemble model. Lastly, Makridakis et al. (2018) are extremely positive about the enormous potential of ML methods for forecasting, noting that these methods have been proposed in the academic literature as alternatives to statistical ones for time series forecasting with scant evidence of their relative performance in terms of accuracy and computational requirements.

After reviewing this literature, we realize that artificial neural networks, to the best of our knowledge, were not applied to interpolate, distribute, and extrapolate time series by related series until now. Table 4 summarizes a comparison between the Chow and Lin (1971) and our approach.

Table 4 - A comparison between two approaches

|  | Chow and Lin | RIDE |
|---|---|---|
| Performance metrics | Minimize error | Minimize error |
| Influenced by the indicator's choice | Yes | Yes |
| Unified approach | Yes | Yes |
| Assumption about variables relationship | Linear | Non-linear |
| Analytic demonstration | Yes | No |
| Big data input | Hard to optimize | Fits well |

## 6.3    Mapping GDP

A relevant task in applied economics is finding metrics that represent the current state of the economy. There is vast information out there, but it is not usually straightforward to isolate the signal from the noise. The official gross domestic product (GDP), this well-known metric customarily disclosed by government statistical institutes worldwide and accompanied by several segments, are usually published quarterly, sometimes yearly, and with lag that can reach several months, given the complexities involved in the calculations of the National Account systems. As a result, many coincident indicators with a wide range of techniques have been developed in economics. The relevance of this variable is indisputable, both for the monetary authority and for diverse economic agents. What is the state of the (...) economy right now? Is it expanding or shrinking, and by how much? These are questions that official GDP statistics try to answer, but they take time to be published. Can we obtain a quicker answer using other data sources? (Hinds et al., 2017, 35). For that purpose, economists have developed models to estimate economic activity in response to the regular ow of data.

Here, we apply deep learning (3.1.1) as a strategy for mapping the Brazilian GDP. With the end-to-end approach allowed by this technique, we propose to map a set of high-frequency variables to fit the GDP. When evaluating an end-to-end strategy, we want to observe whether deep learning models, known as universal approximators, perform well to this task. Since the higher frequency variable is generally not observable, as in the typical case where we want a monthly from a quarterly available GDP, we do not know the actual value to verify the performance, so we adopted two approaches, i) as highlighted in 3.1.3, the factors that determine how well a machine learning algorithm will perform are its ability to make the training error small, and make the gap between training and test error small (Goodfellow et al., 2016, 109), and ii) comparing the results with other approaches, but understanding that none finds the true values.

6.3.1    Implementation

By definition, GDP is the sum of final goods and services produced in an economy over a period, but the access to its components and weights is not straightforward. From the perspective of an end-to-end approach, GDP is a composite index we want to map from their components, even if we have only indirect measures or so-called proxy variables. It should be noted, however, that these proxies are more related to the gross value of production, while we want the added value; for details, see World Bank (2009). In our strategy, the neural network must minimize this problem in its optimization process and because of its non-linear structure, whose performance is measured by the resulting accuracy.

In the task of mapping the Brazilian GDP, we adopted twenty-three monthly variables as indicators, as presented in the Appendix A, Table 9, from January 1996 until March 2021. As mentioned in 6.2, the selection of variables is a relevant point. In the first place, our choice reflects the knowledge about which types of variables are related to GDP. It is an ad hoc decision and follows the logic of two of the most relevant activity indexes for the Brazilian economy, namely, the GDP Monitor from Instituto Brasileiro de Economia (2015), and the Central Bank Economic Activity Index (IBC-Br) from Banco Central do Brasil (2016). These indexes, mainly followed by agents interested in evaluating the current state of the Brazilian economy, adopt the strategy known as accounting indices, that is, an index composed of many proxy variables that are known to be related to the variables used to prepare the official GDP. Naturally, this choice is not perfect, and it encounters additional obstacles. Specifically for our application, we opted for series available for more extended periods, which reduces their quantity. Proxies' variables appear, change, and are discarded over time. If we restrict our sample to a shorter period, we would have more but smaller series. Choosing a longer versus a wider database is a common trade-o in this type of problem.

We computed the input features and the target in level[26] and year over year (YoY) percentage change transformation, both normalized. We split the dataset into train, validation, and test sets. Then, we de ne a function that creates a neural network with hidden layers, ReLU, tanh, or sigmoid as activation functions, and L1 (Lasso) or L2 (Ridge) regularization to reduce overfitting. Next, we retrain the model with the optimal hyperparameters, selected with Hyperband, to evaluate the results, with mean squared error as a loss function and the adaptive moment estimation (Adam) (Kingma and Ba, 2017) for optimization. A browser-based version of the proposed model, without the grid search function, is available at http://www.deeplearningeconomics.com/RIDE/.

6.3.2    Results

The results demonstrate the deep learning's suitability for interpolation, distribution, and extrapolation of time series by related series. Figures 12 to 18 consolidate the results found, that are slightly different each time the models are run due to different compositions in the selected data sets for training, validation, and test. In terms of learning, the curves of the training and validation set behave as expected (Figures 12 and 14), that is, over time (epochs), the highest error reduces and stabilizes at a low level, being the validation error a little higher than the training error.

---

[26] In this case, the resulting monthly GDP is not stationary and presents seasonality. Additional adjustments, commonly required for monthly GDP analyses, can be made a posteriori outside the proposed model.

6.3.2.1   Interpolation and Distribution

The interpolation and distribution concepts (Friedman, 1962) described in 6.2 have evolved in ways that are sometimes taken to be similar, as noted by Issler and Notini (2014)[27]. A subtle difference in the application to GDP data can be seen when some quarters referring to the target (GDP) within the sample are missing, but there are monthly indicators for these periods, in which case it would be necessary to interpolate GDP values, in addition to distributing them in the quarter.

The performance of the deep learning model to generate monthly data is shown in Figures 13 and 15, where we see the good fitting related to the official quarterly data, and in figures 16 and 17 the comparison of RIDE model in level with benchmarks for monthly GDP data, namely the Chow and Lin (1971) method, the GDP Monitor from Instituto Brasileiro de Economia (2015), and the Central Bank Economic Activity Index (IBC-Br)[28] from Banco Central do Brasil (2016). The RIDE correlation (year-over-year growth) with IBC-Br is 0.9266 and with GDP Monitor is 0.9213. Despite using different strategies, the different approaches result in a monthly variable that follows GDP. As we have commented before, none of them can be taken as the authentic higher frequency movement, but they are still essential to guide the decision-making process of governments and the private sector.

---

[27] First, a word of caution about using the term interpolation here. GDP is a ow variable for which we possess quarterly observations we want to distribute within the months in the quarter. [...] the problem of allocating a quarterly ow as such is referred to as distribution, whereas interpolation estimates monthly values of stock variables from quarterly values. Despite this technical distinction for ow and stock variables, several authors still refer to interpolated GDP, a term that is now ingrained in the literature, being the reason why we employ it here (Issler and Notini, 2014, 7).

[28] The IBC-Br presents a level shift in 2015 due to methodological issues.

6.3.2.2   Extrapolation

Lastly, we can use the higher frequency indicator series to infer the GDP already realized but not yet disclosed. There are many questions related to this area of research, one of the most relevant being the difference between near real-time versus real-time proxies[29]. However, these details are not the object of this work. What we can show is the extrapolation capacity of the proposed model. The Figure 18 compares the Brazilian GDP realized between 2020Q3 and 2021Q1 with an out-of-sample RIDE estimate. Like Chow and Lin (1971) and others, RIDE is a global model where the parameters are calculated based on the entire sample; therefore, in cases where the indicator series extends beyond the period cover by the endpoints of the benchmark series, the target will be extrapolated with the global target/indicator ratio. Furthermore, like in the others approaches, when using RIDE the analyst can diversify the use of variables, sample periods and evaluate performance one or several steps ahead, in real or near real-time.

---

[29] Further advice from Kliesen and McCracken (2016) from his work about tracking the US economy is that users of nowcasting models should be aware that most of the monthly source data are initially sample-based estimates. This means, in effect, that the initial estimates are subject to repeated revision as new information becomes available.

Figure 12 - Training and validation curves level

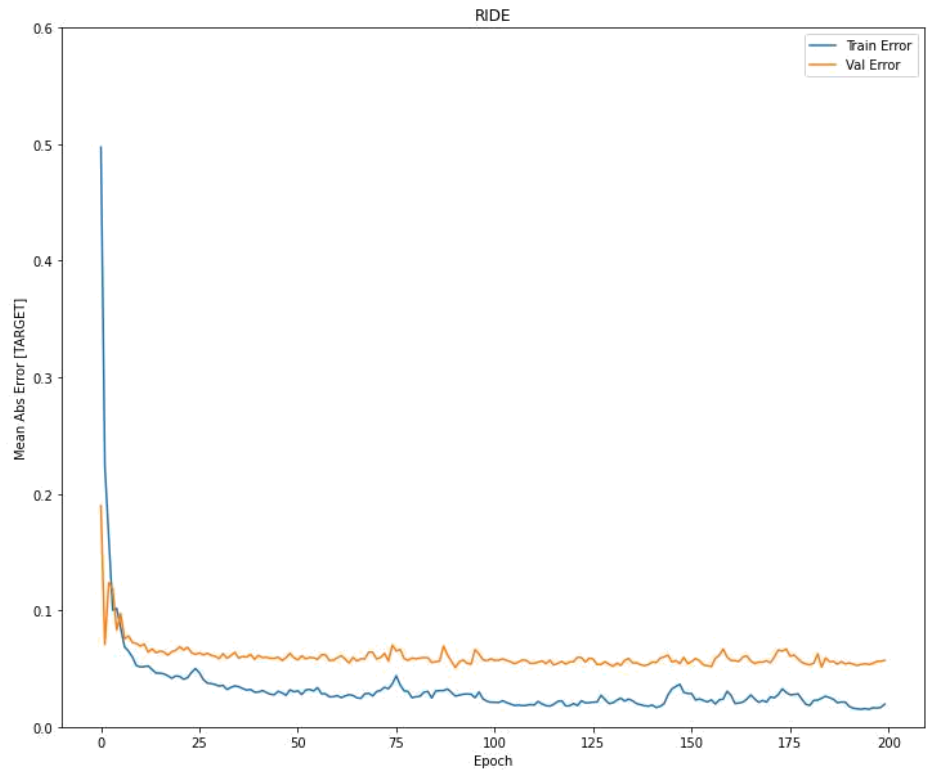

Figure 13 - RIDE mapping GDP level

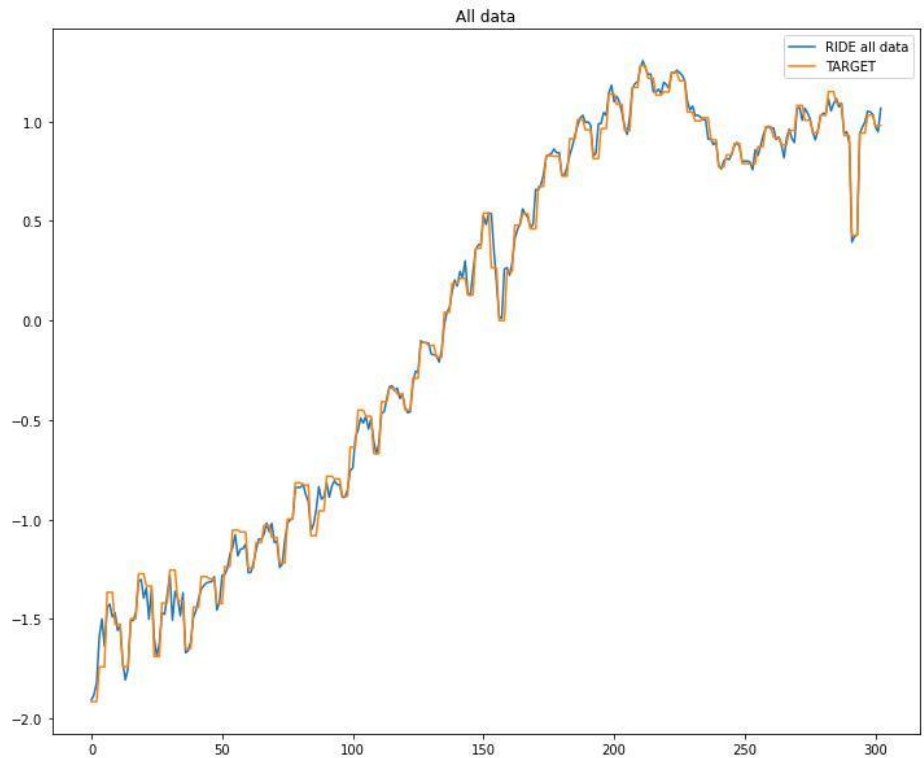

Figure 14 - Training and validation curves YoY

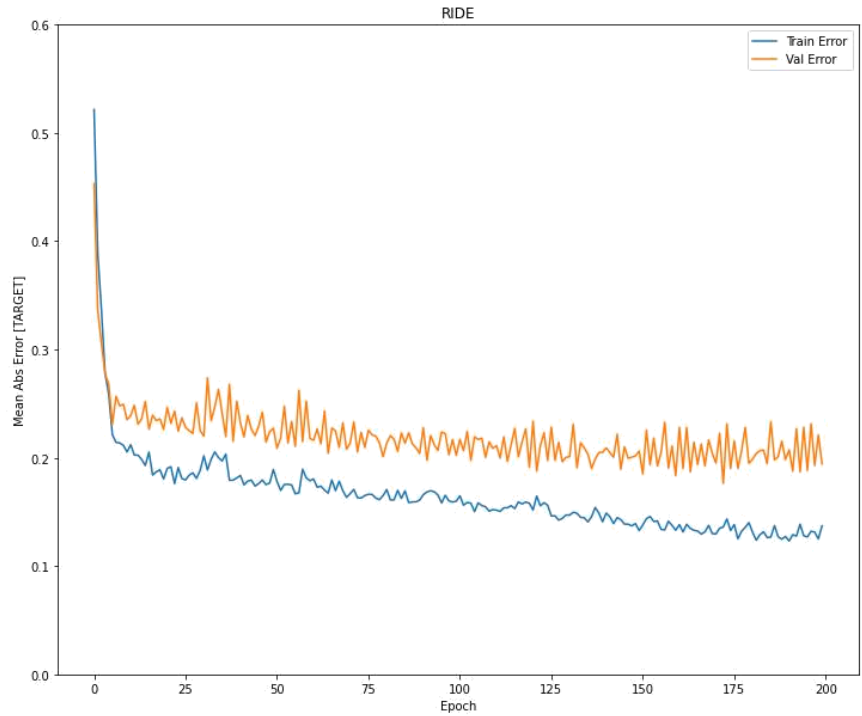

Figure 15 - RIDE mapping GDP YoY

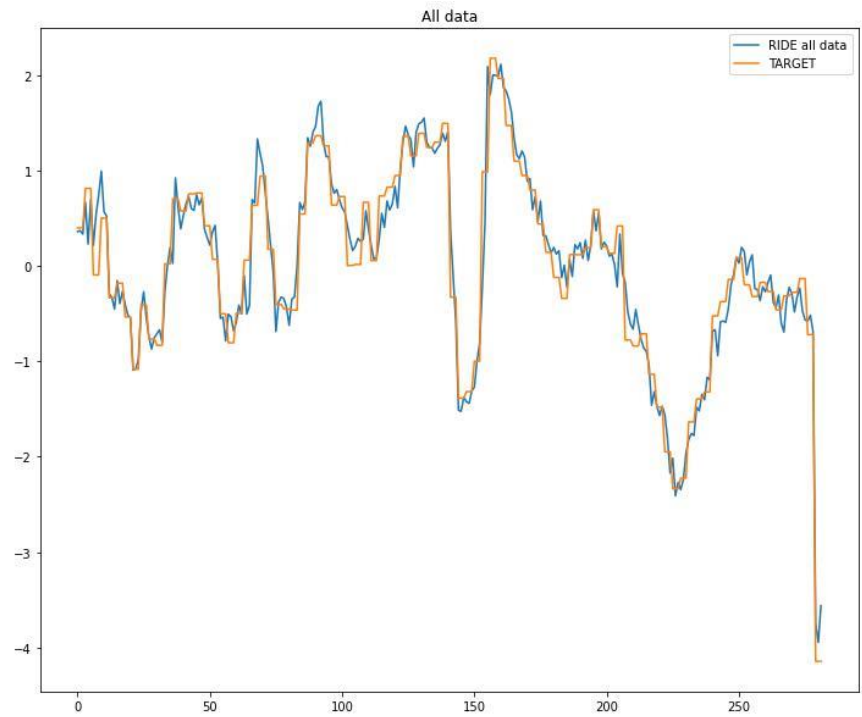

Figure 16 - Chow Lin, IBC-Br and FGV Monitor fitting GDP

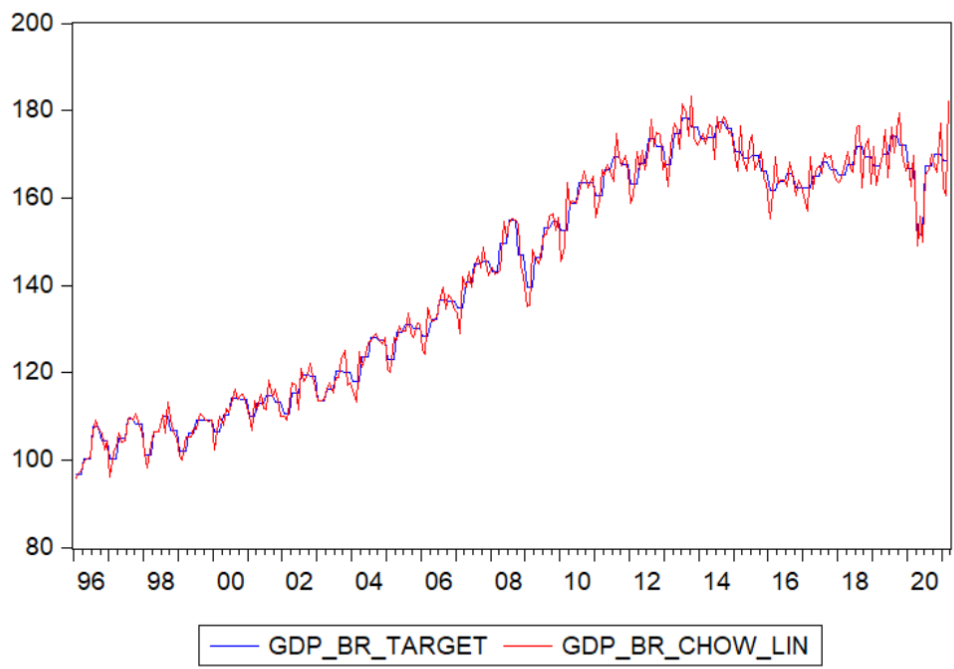

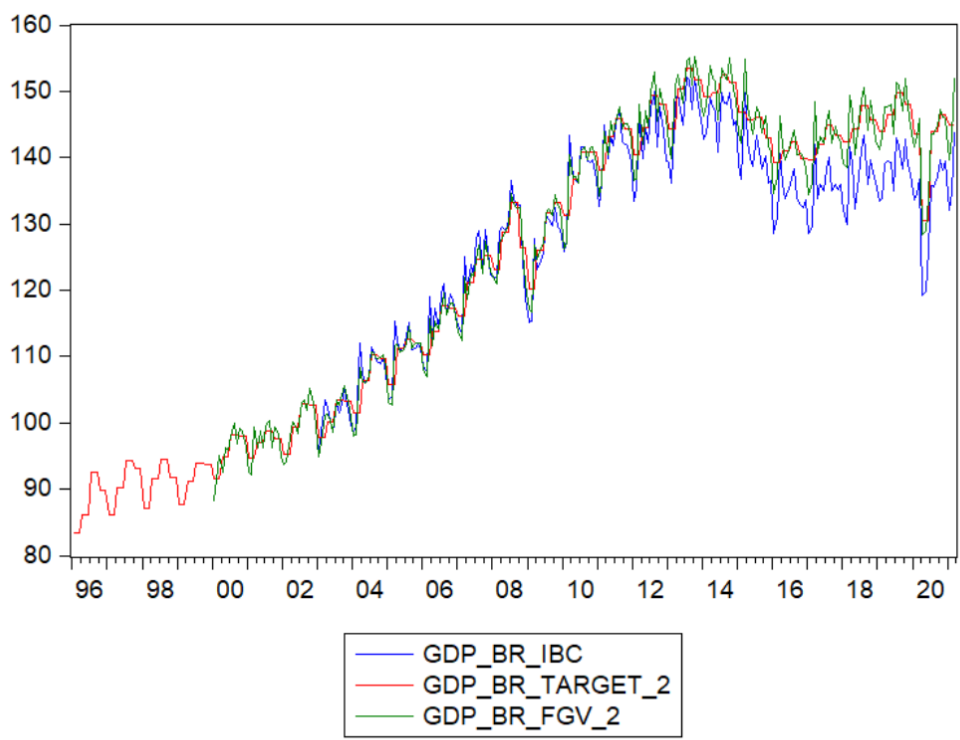

Figure 17 - RIDE, IBC-Br and FGV Monitor

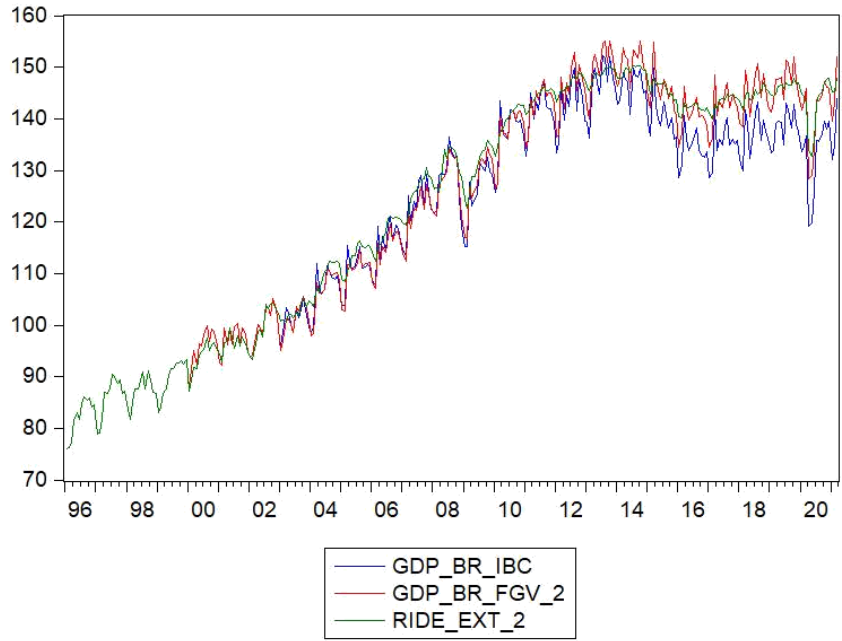

Figure 18 - RIDE extrapolation

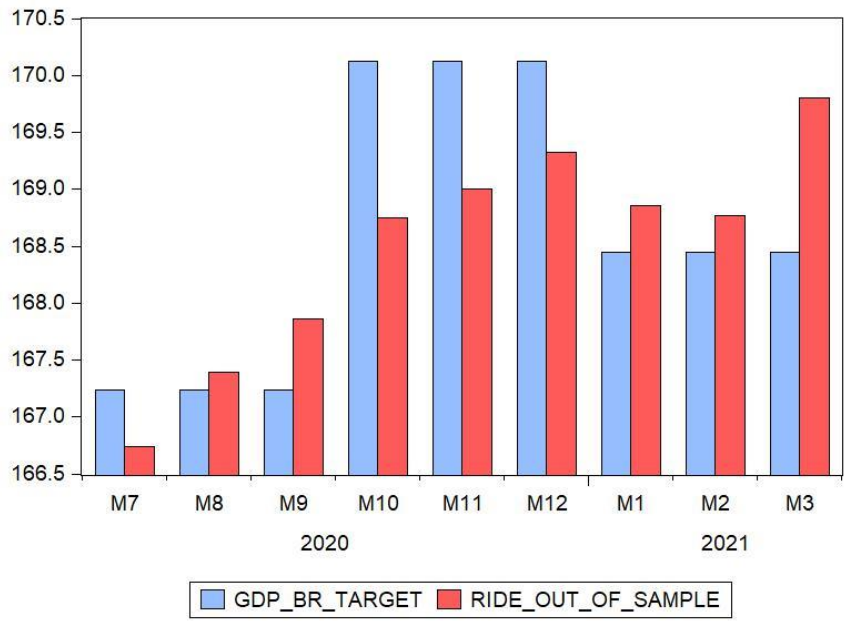

# 7     CONCLUSION

In this work, deep learning was applied to macroeconomics in two approaches: to transfer learning and to interpolate, distribute, and extrapolate time series by related series.

Transfer learning was proposed as an additional strategy for empirical macroeconomics. First, we presented theoretical concepts related to transfer learning and proposed a connection with a typology related to macroeconomic models. At this point, there is already an enormous potential to be explored in the connection between transfer learning and applied macroeconomics. Then, we refer to several algorithms used for transfer learning and detail our choice, artificial neural networks with multiple layers hidden between the input and output layers, also known as deep learning, because it has shown excellent performance for transfer learning, but also because it is significantly less black box than it was in the past. Since feature importance is relevant in economics, neural networks usually are not the first choice. However, interpretability in AI is an active area of research with significant advances, both by researchers' determination and by the requirements of companies and governments to adopt these models for decision-making. Partial Dependence Plots, Permutation Feature Importance, Shapley Value, and Integrated Gradients are available methods to compute feature relevance, to name a few. Secondly, we explore the proposed strategy empirically. As we showed, data from different but related domains, a type of transfer learning, helped to identify the business cycle phases when there was no business cycle dating committee or to estimate a quick first economic-based output gap. In both cases, the proposed strategy also helped to improve the learning when data was limited. The approach demonstrated excellent empirical performance with data from the US, Europe and Brazil, emerging as a potential supplementary tool for governments and the private sector to conduct their activities in the light of national and international economic conditions. The proposed strategy allows a quick application of the economic-based models to other geographic areas. Once trained, the machine inferred the business cycle state and the output gap based on the theory chosen for training. To the best of our knowledge, the combined deep and transfer learning approach is underused for application to macroeconomic problems, indicating that there is plenty of room for research development.

Additionally, since deep learning methods are a way of learning representations, we applied deep learning for mapping low-frequency from high-frequency variables because there are situations where we know, sometimes by construction, that there is a relationship between input and output variables, but this relationship is di cult to map, a challenge in which deep learning models have shown excellent performance, especially towards a data-centric view, where we hold the code fixed and interactively improve the data. The results obtained from Brazilian data demonstrate deep learning's suitability for this task.

As a caveat, it is essential to mention that transfer learning depends not only on properly tuned databases and correctly specified models but also on some critical choices of the econometrician. For example, this work uses data from the American economy for data training and the Brazilian and European economies for transfer. Even though they are regions with their specificities, it can be argued that they are pretty diversified and present relevant similarities. However, suppose that the transfer learning for business cycle classification was applied to a less diversified economy, approaching monoculture. Would the good results persist? Intuitively, it is not expected, which requires that the macroeconomist make different choices, such as using a database for training a source country or region that is more similar to the target region. This fact reinforces the premise that the techniques used here require qualified professionals to take relevant strategic decisions that might affect the results.

Next, this research aims (i) improve the models to enable unstructured data types such as texts and images as inputs, and (ii) in the context of data-centric versus model-centric AI development, move towards the former, building databases to ingest into automated machine learning models.

# APPENDIX A - DATA DESCRIPTION

Figure 19 - U.S. raw data for learning Business Cycle.

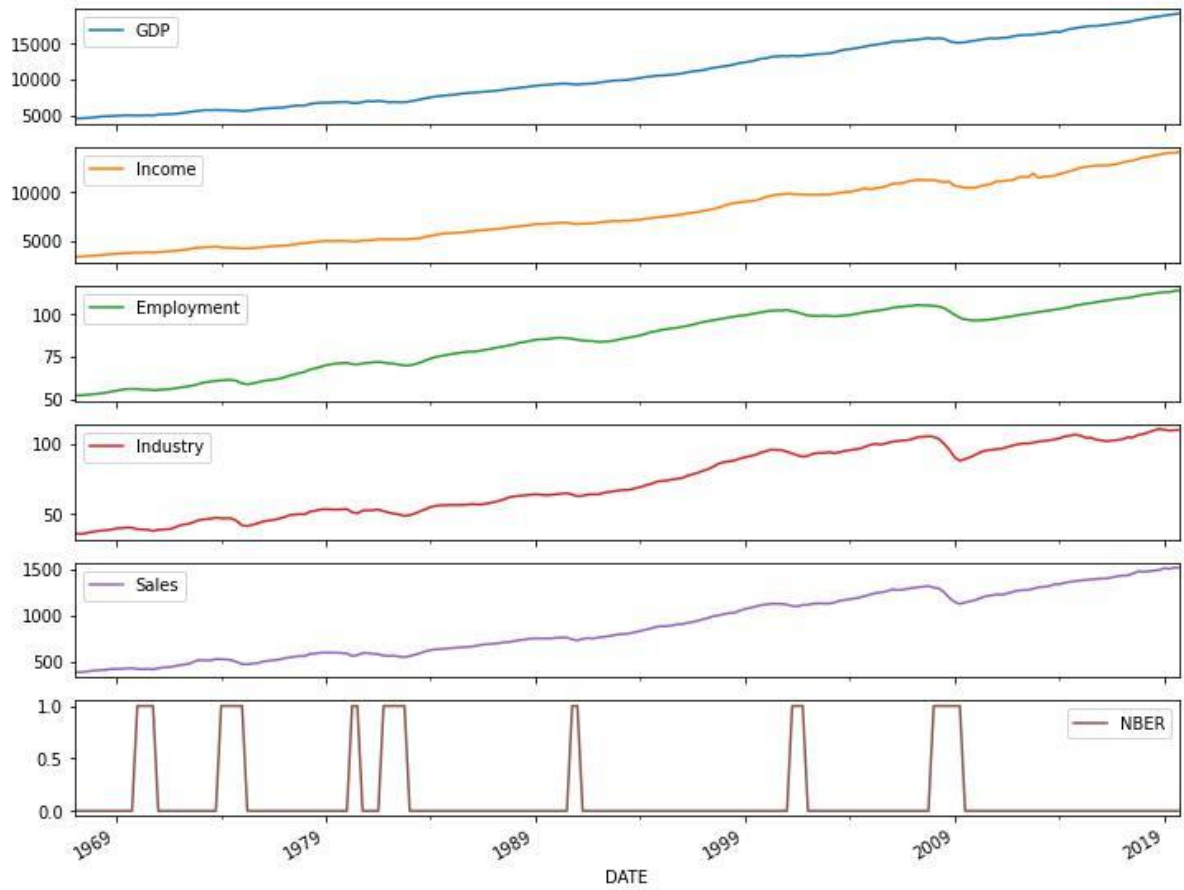

Table 5 - Dataset for Business cycle identification summary

| FRED Code | Series | Start | Characteristics of the raw data | Source |
|---|---|---|---|---|
| | UNITED STATES | | | |
| USRECQ | NBER based Recession Indicator | 1967:Q1 | Recession: 1 = true ; 0 = false | NBER |
| GDPC1 | Real Gross Domestic Product | 1967:Q1 | Billions of Chained 2012 Dollars, s.a. | FRED |
| PIECTR | Real personal income ex current transfers | 1967:Q1 | Billions of Chained 2012 Dollars, s.a. | FRED |
| PRS85006013 | Nonfarm Business Sector employment index | 1967:Q1 | Index 2012 = 100, s.a. | FRED |
| IPB50001SQ | Industrial production index | 1967:Q1 | Index 2012 = 100, s.a. | FRED |
| CQRMTSPL | Real manufacturing and trade ind. sales | 1967:Q1 | Millions of Chained 2012 Dollars, s.a., | FRED |
| | EURO AREA | | | |
| N/A | CEPR based Recession Indicator | 2005:Q1 | Recession: 1 = true \| 0 = false | CEPR |
| CLVMNACSCAB1GQEA19 | Real Gross Domestic Product (19 countries) | 2005:Q1 | Millions of Chained 2010 Euros, s.a. | FRED |
| NAEXKP02EZQ189S | Private Final Consumption Expenditure | 2005:Q1 | Billions of Chained 2012 Dollars, s.a. | FRED |
| LFESEETTEZQ647S | Employees | 2005:Q1 | Persons, s.a. | FRED |
| PRMNTO01EZQ657S | Total Manufacturing Production | 2005:Q1 | Growth Rate Previous Period, s.a. | FRED |
| SLRTTO01EZQ657S | Volume of Total Retail Trade sales | 2005:Q1 | Growth Rate Previous Period | FRED |
| | BRAZIL | | | |
| N/A | CODACE based Recession Indicator | 2000:Q1 | Recession: 1 = true \| 0 = false | CODACE |
| NAEXKP01BRQ652S | Total Gross Domestic Product | 2000:Q1 | Chained 2000 Real, s.a. | FRED |
| NAEXKP02BRQ189S | Private Final Consumption Expenditure | 2000:Q1 | Chained 2000 Real, s.a. | FRED |
| N/A | Registered Employees Index | 2000:Q1 | Index Dez-2001 = 100, s.a. | BCB |
| BRAPROINDQISMEI | Production of Total Industry | 2000:Q1 | Index 2015 = 100, s.a. | FRED |
| BRASARTQISMEI | Total Retail Trade | 2000:Q1 | Index 2015 =100, s.a. | FRED |

Table 6 - Dataset for Business cycle identification descriptive statistics

| Series | count | mean | std | min | max |
|---|---|---|---|---|---|
| UNITED STATES | | | | | |
| NBER based Recession Indicator | 211 | 0.127962 | 0.334842 | 0.000000 | 1.000000 |
| Real Gross Domestic Product | 211 | 0.006844 | 0.007829 | -0.021876 | 0.037915 |
| Real personal income ex current transfers | 211 | 0.006885 | 0.009104 | -0.039786 | 0.033266 |
| Nonfarm Business Sector employment index | 211 | 0.003678 | 0.006319 | -0.023668 | 0.019575 |
| Industrial production index | 211 | 0.005290 | 0.014928 | -0.068358 | 0.041548 |
| Real manufacturing and trade ind. sales | 211 | 0.006486 | 0.013956 | -0.049767 | 0.049969 |
| EURO AREA | | | | | |
| CEPR based Recession Indicator | 59 | 0.183333 | 0.390205 | 0.000000 | 1.000000 |
| Real Gross Domestic Product (19 countries) | 59 | 0.002860 | 0.006591 | -0.032153 | 0.011921 |
| Private Final Consumption Expenditure | 59 | 0.002141 | 0.003526 | -0.005379 | 0.009745 |
| Employees | 59 | 0.002086 | 0.003732 | -0.010767 | 0.007213 |
| Total Manufacturing Production | 59 | 0.001266 | 0.021143 | -0.108820 | 0.033571 |
| Volume of Total Retail Trade sales | 59 | 0.001792 | 0.005575 | -0.018082 | 0.010880 |
| BRAZIL | | | | | |
| CODACE based Recession Indicator | 79 | 0.225000 | 0.420217 | 0.000000 | 1.000000 |
| Total Gross Domestic Product | 79 | 0.005596 | 0.011576 | -0.040186 | 0.024456 |
| Private Final Consumption Expenditure | 79 | 0.006708 | 0.011400 | -0.031874 | 0.028647 |
| Registered Employees Index | 79 | 0.007989 | 0.008475 | -0.012135 | 0.024315 |
| Production of Total Industry | 79 | 0.002106 | 0.023298 | -0.101341 | 0.050569 |
| Total Retail Trade | 79 | 0.008367 | 0.015890 | -0.033902 | 0.039346 |

Figure 20 - U.S. raw data for learning Output Gap

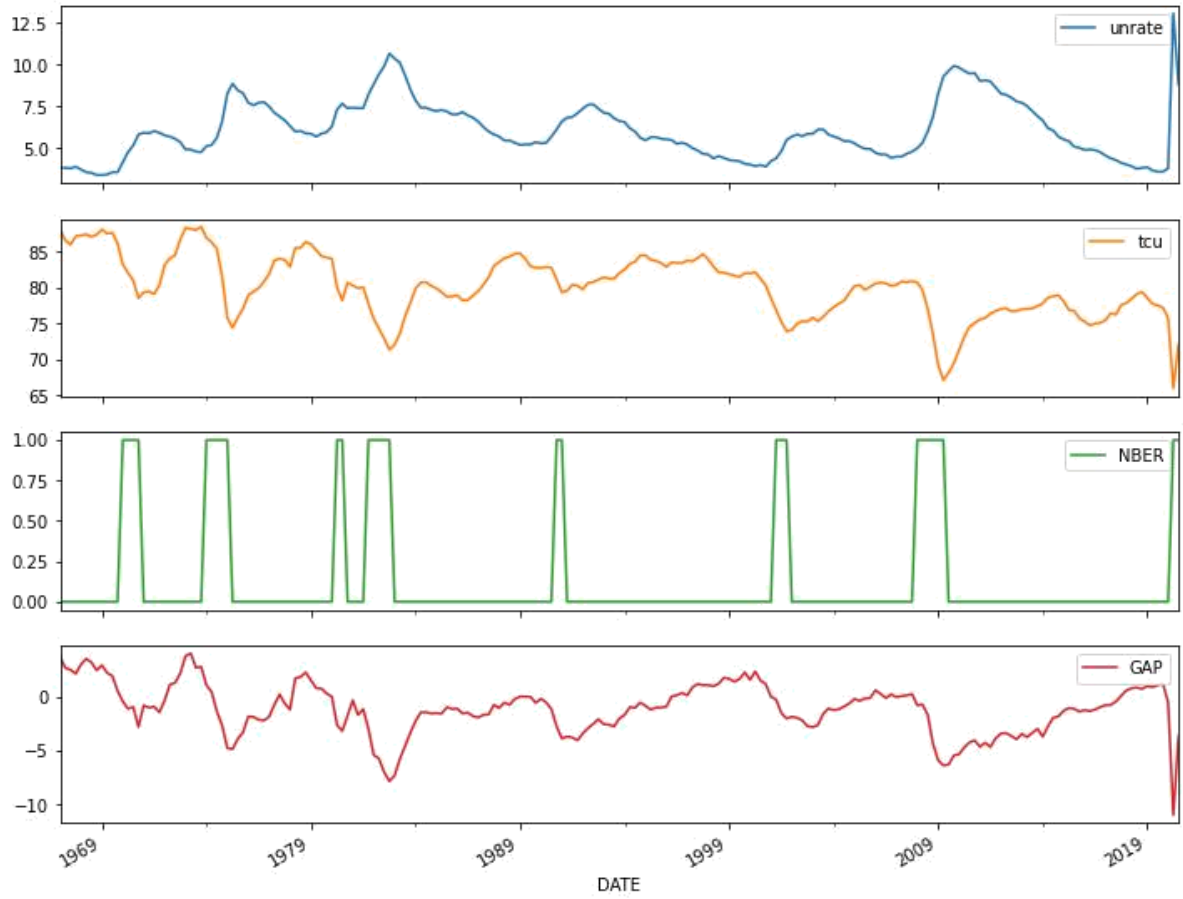

Table 7 - Dataset for Output Gap Estimation: summary

| FRED Code | Series | Start | Characteristics of the raw data | Source |
|---|---|---|---|---|
| | UNITED STATES | | | |
| USRECQ | NBER based Recession Indicator | 1967:Q1 | Recession: 1 = true ; 0 = false | NBER |
| GDPC1 | Real Gross Domestic Product | 1967:Q1 | Billions of Chained 2012 Dollars, s.a. | FRED |
| GDPPOT | Real potential Gross Domestic Product | 1967:Q1 | Billions of Chained 2012 Dollars | FRED |
| UNRATE | Unemployment rate | 1967:Q1 | percent, s.a., quarterly average | FRED |
| TCU | Capacity utilization index | 1967:Q1 | percent, s.a., quarterly average | FRED |
| | BRAZIL | | | |
| N/A | CODACE based Recession Indicator | 2012:Q2 | Recession: 1 = true \| 0 = false | CODACE |
| NAEXKP01BR | Total Gross Domestic Product | 2012:Q2 | Chained 2000 Real, s.a. | FRED |
| TFP | Total Factor Productivity Level (PPP Brazil and U.S.) | 2012:Q2 | Index U.S. = 1 | FRED |
| N/A | Registered Employees Index | 2000:Q1 | Index Dez-2001 = 100, s.a. | BCB |

Table 8  Dataset for Output Gap Estimation: descriptive statistics

| Series | count | mean | std | min | max |
|---|---|---|---|---|---|
| UNITED STATES | | | | | |
| Output GAP | 215 | -1.182651 | 2.345025 | -10.990000 | 3.980000 |
| NBER Indicator | 215 | 0.134884 | 0.342397 | 0.000000 | 1.000000 |
| Unemployment rate | 215 | 6.091907 | 1.730741 | 3.400000 | 13.070000 |
| Capacity utilization index | 215 | 80.094558 | 4.292762 | 65.970000 | 88.530000 |
| BRAZIL | | | | | |
| Output GAP | 34 | -1.026471 | 2.316385 | -4.900000 | 2.400000 |
| CODACE Indicator | 34 | 0.411765 | 0.499554 | 0.000000 | 1.000000 |
| Capacity utilization index | 34 | 76.961765 | 4.366775 | 61.400000 | 82.730000 |
| Total Factor Productivity | 34 | 0.544765 | 0.037840 | 0.506000 | 0.608000 |
| Registered Employees Index | 34 | 10.074118 | 2.557549 | 6.470000 | 14.270000 |

Table 9  Dataset for RIDE: summary

| Code | Series | Range | Source |
|---|---|---|---|
| gdp | Gross domestic product - Target | 1996:M1 to 2021:M3 | IBGE |
| epeRes | Electricity consumption: household | 1996:M1 to 2021:M3 | Empresa de Pesquisa Energetica |
| epeCom | Electricity consumption: commerce | 1996:M1 to 2021:M3 | Empresa de Pesquisa Energetica |
| epeInd | Electricity consumption: industry | 1996:M1 to 2021:M3 | Empresa de Pesquisa Energetica |
| expQ | Export goods: quantum | 1996:M1 to 2021:M3 | Funcex |
| impQ | Import good: quantum | 1996:M1 to 2021:M3 | Funcex |
| expBasic | Export goods: raw material | 1996:M1 to 2021:M3 | Funcex |
| expSemi | Export goods: semi | 1996:M1 to 2021:M3 | Funcex |
| expManuf | Export goods: industrial | 1996:M1 to 2021:M3 | Funcex |
| expAgroDolar | Export goods: agro | 1996:M1 to 2021:M3 | Funcex |
| anfaveaProd | Vehicles production | 1996:M1 to 2021:M3 | IPEA data |
| anfaveaLic | Vehicles sales | 1996:M1 to 2021:M3 | IPEA data |
| cagedTransf | Formal jobs: industry | 1996:M1 to 2021:M3 | Ministry of Labor |
| cagedConstr | Formal jobs: construction | 1996:M1 to 2021:M3 | Ministry of Labor |
| cagedCom | Formal jobs: commerce | 1996:M1 to 2021:M3 | Ministry of Labor |
| cagedServ | Formal jobs: services | 1996:M1 to 2021:M3 | Ministry of Labor |
| ipa | Industrial index prices | 1996:M1 to 2021:M3 | IPEA data |
| cambio | Exchange rate | 1996:M1 to 2021:M3 | IPEA data |
| poup | Savings | 1996:M1 to 2021:M3 | BCB |
| m1 | M1 money supply | 1996:M1 to 2021:M3 | BCB |
| m2 | M2 money supply | 1996:M1 to 2021:M3 | BCB |
| ibov | Stock market | 1996:M1 to 2021:M3 | IPEA data |
| embi | Sovereign risk | 1996:M1 to 2021:M3 | IPEA data |
| RecRFBIPCA | Taxes | 1996:M1 to 2021:M3 | Receita Federal do Brasil |

Table 10  Dataset for RIDE: level descriptive statistics

| Series | count | mean | std | min | max |
|---|---|---|---|---|---|
| Gross domestic product | 294 | 4.932874 | 0.190581 | 4.573057 | 5.184117 |
| Electricity: household | 294 | 9.011303 | 0.248722 | 8.532279 | 9.465525 |
| Electricity: commerce | 294 | 8.535415 | 0.315680 | 7.881182 | 9.012377 |
| Electricity: industry | 294 | 9.456611 | 0.160437 | 9.077152 | 9.673193 |
| Export goods: quantum | 294 | 4.433535 | 0.384342 | 3.443938 | 4.974524 |
| Import good: quantum | 294 | 4.735525 | 0.382289 | 3.694116 | 5.403533 |
| Export goods: raw material | 294 | 4.544279 | 0.605518 | 2.989211 | 5.560105 |
| Export goods: semi | 294 | 4.510472 | 0.294411 | 3.706719 | 4.994912 |
| Export goods: industrial | 294 | 4.273739 | 0.306697 | 3.439135 | 4.718231 |
| Export goods: agro | 294 | 6.975650 | 0.979179 | 4.695358 | 8.618163 |
| Vehicles production | 294 | 12.148554 | 0.446944 | 7.521318 | 12.772318 |
| Vehicles sales | 294 | 12.108557 | 0.386983 | 10.669234 | 12.948200 |
| Formal jobs: industry | 294 | 8.810780 | 0.121875 | 8.586517 | 9.002094 |
| Formal jobs: construction | 294 | 7.724365 | 0.207142 | 7.452128 | 8.112908 |
| Formal jobs: commerce | 294 | 8.839708 | 0.265457 | 8.452849 | 9.162869 |
| Formal jobs: services | 294 | 9.547007 | 0.230508 | 9.233353 | 9.829924 |
| Industrial index prices | 294 | 5.635644 | 0.554824 | 4.603969 | 6.465277 |
| Exchange rate | 294 | 0.80994 | 0.388919 | -0.021632 | 1.700375 |
| Savings | 294 | 19.362327 | 0.796986 | 17.978074 | 20.669209 |
| M1 money supply | 294 | 18.797382 | 0.822260 | 16.940591 | 20.057678 |
| M2 money supply | 294 | 20.502025 | 0.932023 | 18.964322 | 22.002361 |
| Stock market | 294 | 10.480367 | 0.752160 | 8.762259 | 11.661035 |
| Sovereign risk | 294 | 5.940150 | 0.622108 | 4.955827 | 7.781139 |
| Taxes | 294 | 11.398053 | 0.360862 | 10.569928 | 12.011231 |

Table 11  Dataset for RIDE: year over year descriptive statistics

| Series | count | mean | std | min | max |
|---|---|---|---|---|---|
| Gross domestic product | 282 | 0.021008 | 0.032664 | -0.114355 | 0.092091 |
| Electricity: household | 282 | 0.033728 | 0.065733 | -0.272383 | 0.0196253 |
| Electricity: commerce | 282 | 0.041736 | 0.067517 | -0.252346 | 0.230900 |
| Electricity: industry | 282 | 0.016326 | 0.067701 | -0.169337 | 0.242843 |
| Export goods: quantum | 282 | 0.056994 | 0.116426 | -0.257154 | 0.436872 |
| Import good: quantum | 282 | 0.054362 | 0.179768 | -0.319535 | 0.904999 |
| Export goods: raw material | 282 | 0.097804 | 0.174419 | -0.393541 | 0.891475 |
| Export goods: semi | 282 | 0.044599 | 0.144330 | -0.254536 | 0.568317 |
| Export goods: industrial | 282 | 0.040478 | 0.154273 | -0.382852 | 0.493096 |
| Export goods: agro | 282 | 0.171863 | 0.367463 | -0.510009 | 1.630786 |
| Vehicles production | 282 | 0.037177 | 0.236936 | -0.993097 | 1.634799 |
| Vehicles sales | 282 | 0.034511 | 0.228297 | -0.759697 | 1.617758 |
| Formal jobs: industry | 282 | 0.006597 | 0.036823 | -0.085214 | 0.086855 |
| Formal jobs: construction | 282 | 0.005617 | 0.070574 | -0.150847 | 0.147416 |
| Formal jobs: commerce | 282 | 0.028729 | 0.028792 | -0.033498 | 0.072822 |
| Formal jobs: services | 282 | 0.024881 | 0.027269 | -0.024992 | 0.108166 |
| Industrial index prices | 282 | 0.081762 | 0.073732 | -0.051507 | 0.403623 |
| Exchange rate | 282 | 0.086137 | 0.216944 | -0.254942 | 0.826610 |
| Savings | 282 | 0.118122 | 0.081935 | -0.25782 | 0.390536 |
| M1 money supply | 282 | 0.135416 | 0.126388 | -0.052146 | 0.695404 |
| M2 money supply | 282 | 0.132639 | 0.082459 | -0.003356 | 0.430383 |
| Stock market | 282 | 0.108480 | 0.371873 | -0.881164 | 1.414691 |
| Sovereign risk | 282 | 0.059905 | 0.499091 | -0.708559 | 2.683333 |
| Taxes | 282 | 0.044488 | 0.091628 | -0.325202 | 0.367138 |

**APPENDIX B - SELECTED CODES**

```python
neg, pos = np.bincount(raw_data['NBER'])
total = neg + pos
# set initial bias
initial_bias = np.log([pos/neg])

# Use a utility from sklearn to split out dataset
train_df, test_df = train_test_split(log_1df, test_size=0.4, random_state=0,
shuffle=False)
train_df, val_df = train_test_split(train_1df, test_size=0.3, random_state=0,
shuffle=False)
# Form np arrays of labels and features
train_labels = np.array(train_df.pop('NBER'))
val_labels = np.array(val_df.pop('NBER'))
test_labels = np.array(test_df.pop('NBER'))

train_features = np.array(train_df)
val_features = np.array(val_df)
test_features = np.array(test_df)

scaler = StandardScaler()
train_features = scaler.fit_transform(train_features)
val_features = scaler.transform(val_features)
test_features = scaler.transform(test_features)

METRICS = [
    keras.metrics.TruePositives(name='tp'),
    keras.metrics.FalsePositives(name='fp'),
    keras.metrics.TrueNegatives(name='tn'),
    keras.metrics.FalseNegatives(name='fn'),
    keras.metrics.BinaryAccuracy(name='accuracy'),
    keras.metrics.Precision(name='precision'),
    keras.metrics.Recall(name='recall'),
    keras.metrics.AUC(name='auc'),
]

def make_model(hp, metrics = METRICS, output_bias = initial_bias):
    output_bias = tf.keras.initializers.Constant(output_bias)

    model = keras.Sequential()

    # Tune the number of units in the layers
    # Choose an optimal value between 16-256
    hp_units = hp.Int('units', min_value = 16, max_value = 256, step = 16)

    model.add(keras.layers.LSTM(units = hp_units, input_shape =
(1,train_features.shape[-1],), dropout = 0.3)) #LSTM layer
```

```python
        model.add(keras.layers.Dense(units = hp_units, activation = 'relu'))
#Dense1
        model.add(keras.layers.Dense(units = hp_units, activation = 'relu'))
#Dense2
        model.add(keras.layers.Dense(units = hp_units, activation = 'relu'))
#Dense3
        model.add(keras.layers.Dense(units = hp_units, activation = 'relu'))
#Dense4

        model.add(keras.layers.Dropout(0.5)) # To prevent overfiting

        model.add(keras.layers.Dense(1, activation='sigmoid',
bias_initializer=output_bias)) # Output layer

        # Tune the learning rate for the optimizer
        # Choose an optimal value from 0.01, 0.001, or 0.0001
        hp_learning_rate = hp.Choice('learning_rate', values = [1e-2, 1e-3, 1e-
4])

        model.compile(optimizer = keras.optimizers.Adam(learning_rate =
hp_learning_rate),
            loss = keras.losses.BinaryCrossentropy(),
            metrics = metrics)

    return model

tuner = kt.Hyperband(make_model,
                     kt.Objective('val_auc', direction='max'), #maximizes area
under the ROC curve
                     max_epochs = 10,
                     factor = 3,)

tuner.search(train_features, train_labels,
             epochs=50,
             validation_data=(val_features, val_labels), callbacks =
[ClearTrainingOutput()])

#Get the optimal hyperparameters
best_hps = tuner.get_best_hyperparameters(num_trials = 1)[0]

#Retrain the model with the optimal hyperparameters
model = tuner.hypermodel.build(best_hps)
baseline_history = model.fit(train_features, train_labels,
                             epochs=50,
                             validation_data=(val_features, val_labels))
```